\documentclass[12pt, draftclsnofoot, onecolumn]{IEEEtran}
\usepackage{graphicx}
\usepackage{amsmath,bbm,epsfig,amssymb,amsfonts, amstext, verbatim,amsopn,multirow,multicol,lipsum}

\usepackage[font=scriptsize]{caption}
\usepackage{subcaption}
\usepackage{balance}
\usepackage{url}
\usepackage{bbding}
\usepackage{amsfonts}
\usepackage{epsfig}
\usepackage{epstopdf}
\usepackage{setspace}
\usepackage{stmaryrd}
\usepackage{psfrag}	
\usepackage{multirow}
\usepackage{float}
\usepackage{cases}  
\usepackage[process=auto]{pstool}
\usepackage{etoolbox}
\usepackage{algorithm}
\usepackage{algorithmic}
\usepackage{hyperref}
\usepackage{wasysym}[nointegrals]
\usepackage{pifont}
\usepackage{enumitem}
\allowdisplaybreaks
\usepackage{graphicx}
\usepackage{wrapfig}
\usepackage{lscape}
\usepackage{rotating}
\usepackage{epstopdf}

\newtheorem{theo}{Theorem}
\newtheorem{lem}{Lemma}

\newtheorem{prop}{Proposition}
\newtheorem{corol}{Corollary}

\newcommand{\pushright}[1]{\ifmeasuring@#1\else\omit\hfill$\displaystyle#1$\fi\ignorespaces}
\newcommand{\pushleft}[1]{\ifmeasuring@#1\else\omit$\displaystyle#1$\hfill\fi\ignorespaces}
\newtoggle{Conf}

\togglefalse{Conf}

\iftoggle{Conf}{%
}{%
}

\EndPreamble

\usepackage{rotating}

\begin{document}
\title{ Optical IRSs: Power Scaling Law, Optimal Deployment, and Comparison with Relays 
\vspace{-0.3cm}\footnote{ This paper was presented in part at IEEE Globecom 2022 \cite{Globecom_ScalingLaw}.}}
\author{Hedieh Ajam\IEEEauthorrefmark{1}, Marzieh Najafi\IEEEauthorrefmark{1}, Vahid Jamali\IEEEauthorrefmark{2},   and Robert Schober\IEEEauthorrefmark{1}\\  
		\IEEEauthorrefmark{1}Friedrich-Alexander-Universi\"at  Erlangen-N\"urnberg, \IEEEauthorrefmark{2}Technical University of Darmstadt
		\vspace{-0.4cm}}
	\maketitle
\begin{abstract}
	The line-of-sight (LOS) requirement of free-space optical (FSO) systems can be relaxed by employing optical relays or optical intelligent reflecting surfaces (IRSs). 	 In this paper, we show that  the   power  reflected from FSO IRSs and collected at the receiver (Rx) lens  may scale quadratically or  linearly  with the IRS size or  may   saturate at a  constant value. We analyze the power scaling law for optical IRSs and unveil its dependence  on the wavelength,  transmitter (Tx)-to-IRS and  IRS-to-Rx distances,  beam waist, and Rx lens size. We also consider the impact of linear, quadratic, and focusing phase shift profiles across the IRS on the power collected at the Rx lens for different IRS sizes. Our results reveal that surprisingly  the  powers received for the different phase shift  profiles are identical, unless the IRS operates in the saturation regime. Moreover,  IRSs employing the focusing (linear) phase shift profile  require the largest (smallest) size to reach the saturation regime.	 	We also compare optical IRSs in different power scaling regimes with  optical relays  in terms of the outage probability,  diversity and  coding gains, and  optimal placement. Our results show that, at the expense of a  higher hardware complexity, relay-assisted FSO links yield a better outage performance at  high signal-to-noise-ratios (SNRs), but  optical IRSs can achieve a higher  performance at low SNRs. Moreover, while it  is optimal to place relays equidistant from Tx and Rx, the optimal location of  optical IRSs  depends on the  phase shift profile and the power scaling regime  they operate in.  
\end{abstract}
	\section{Introduction}
Due to their  directional narrow laser beams and easy-to-install transceivers, free space optical (FSO) systems   are promising candidates for  high data rate applications, such as wireless front- and backhauling,  in  next generation  wireless communication networks and beyond \cite{6G}.   FSO systems  require a line-of-sight (LOS) connection between transmitter (Tx) and receiver (Rx)  which can be relaxed by using  optical relays \cite{Safari_relay} or optical intelligent reflecting surfaces (IRSs) \cite{ Marzieh_IRS_jou, AmplifyingRIS_Herald}. Optical relays can either process and decode the received signal or only amplify it before retransmitting it  to the Rx. For high data rate FSO systems, relays may require high-speed decoding and encoding hardware and/or analog gain units, additional synchronization, and clock recovery \cite{All-optical}.  On the other hand, optical IRSs are planar structures comprised of passive subwavelength elements, known as unit cells, which  can manipulate the  properties of an incident wave such as its phase and polarization \cite{ Marco_survey,IRS-FSO-Herald}. In particular, to redirect an incident beam in a desired direction, the IRS can  apply  a phase shift to the incident wave and thereby adjust the accumulated phase of the reflected wave \cite{Marzieh_IRS, TCOM_IRSFSO}. Optical IRSs can be implemented using mirror- or metamaterial (MM)-based  technologies \cite{Vahid_Magazine}  and, due to their passivity, incur a lower  system complexity and power consumption than optical relays. 

For  radio frequency (RF) IRSs, by increasing the IRS area, $\Sigma_\text{irs}$,  the received power  scales quadratically  \cite{RuiZhang_PowerScale,Emil-PowerScale}.  For very large IRS sizes, the received power saturates  at $\frac{1}{9}$ of the transmit power because as the IRS gets larger, its relative effective area (area perpendicular to the propagation direction) gets smaller \cite{Emil-PowerScale}.   However, there is a fundamental difference between FSO systems and the RF systems considered in \cite{Emil-PowerScale}.    In \cite{Emil-PowerScale}, the Rx area is smaller than a wavelength whereas  optical lenses are typically much larger than a wavelength.  Moreover,    spherical/planar RF waves  lead to a uniform power distribution across the IRS, whereas FSO systems employ  Gaussian laser beams, which have a curved wavefront and a non-uniform power distribution \cite{TCOM_IRSFSO}. Furthermore, the electrical size of the IRS (IRS length divided by the wavelength) at optical frequencies is much larger than at RF. Thus, we expect the power scaling law for optical IRSs to differ from that of RF IRSs. However, except for the preliminary results reported in the conference version of this paper \cite{Globecom_ScalingLaw}, the power scaling law of optical IRSs has not been investigated yet. 

In this  paper, we analyze the power scaling law for optical IRSs in detail and show that  depending on the Rx lens size,  the locations of  Tx and Rx with respect to (w.r.t.) the IRS, and the beam waist,  the  received power may scale quadratically ($\mathcal{O}(\Sigma_\text{irs}^2)$) or linearly ($\mathcal{O}(\Sigma_\text{irs})$) with the  IRS size or it may  saturate to a constant value ($\mathcal{O}(1)$). We will show that when both IRS and lens are smaller than  footprint of the incident beam, the received power scales quadratically with the IRS size ($\mathcal{O}(\Sigma_\text{irs}^2)$).  On the other hand, when the lens is larger than the  beam footprint but the IRS is still smaller than the  beam footprint, the  received power scales linearly with the IRS size ($\mathcal{O}(\Sigma_\text{irs})$).  Finally, if  both IRS and lens  are  larger than the   footprint of the incident beam, the received power saturates at a constant value ($\mathcal{O}(1)$).

Furthermore, we investigate the impact of different phase shift profiles of  MM-based IRSs on the power scaling law. We show that for linear (LP), quadratic (QP), and focusing  (FP) phase shift profiles   the received powers at the lens  are identical when the IRS operates in the quadratic and linear power scaling regimes. However, the adopted  phase shift profile does affect the received power when the IRS  operates in the saturation regime. In this case,  the FP profile yields the largest received power as it focuses the total incident beam on the lens center, whereas the LP and QP profiles yield smaller  values. In addition, we show that  mirror-based IRSs  outperform   MM-based IRSs with  LP profile due to their ability  to adjust their orientation to the incident beam.

We also compare  the performance of relay- and IRS-assisted FSO systems. Such comparisons were made between  RF IRSs and  decode-and-forward (DF) and amplify-and-forward (AF) relays in \cite{IRSvsRel_Larsson} and \cite{IRSvsAF_Debbah}, respectively.  However, RF links are fundamentally different from FSO links, where the variance of the fading  is distance-dependent \cite{Safari_relay}.   Moreover, to reduce hardware complexity,  often half-duplex relays are  preferred for RF systems,  whereas FSO relays are typically full-duplex \cite{Sahar_Placement}.

In this paper, we study relay- and IRS-assisted FSO systems where  our contributions can be summarized as follows:
\begin{itemize}
\item We analyze the power scaling law for different  IRS   and Rx lens sizes,  and show that depending on the size of the beam footprint on the IRS  and  the lens, the received power may scale  linearly or quadratically  with the IRS size or it may remain constant.
\item We consider different phase shift profiles for  MM-based IRSs and show that the impact of the phase profile is  negligible when the IRS  size is much smaller than the incident beam footprint on the IRS. However, when the IRS size is comparable to the beam footprint on the IRS,   IRSs with QP and FP profiles perform better than IRSs with LP profile. 
\item We compare the performance of   IRS- and  relay-assisted  FSO systems  in terms of their outage probability and their diversity and coding gains. Our results  show that, at the expense of a higher hardware complexity,  relay-assisted FSO links yield a higher  diversity gain as the variance of the corresponding distance-dependent fading is smaller compared to that for  IRS-assisted FSO links.  Moreover, the coding gain in IRS-assisted FSO links may increase with the IRS size depending on the power scaling regime the IRS operates in.
\item We also analyze the  optimal positions of  FSO relays and IRSs for minimization of  the end-to-end  outage performance at high SNRs. We show that while  relays are optimally  positioned  equidistant from   Tx and Rx \cite{Safari_Placement, Sahar_Placement}, the optimal position of optical  IRSs depends on the  power scaling regime  they operate in as well as the IRS technology and phase shift profile. In particular, at the expense of a large form factor needed for rotation, mirror-based IRSs perform identical at any position between Tx and Rx. In contrast,   MM-based IRSs with LP profile achieve optimal outage performance   close to Tx, close to Tx or Rx,  and  equidistant  from Tx and Rx when they operate in the linear, quadratic,  and saturation power scaling regime, respectively. Furthermore,  for IRSs operating in the  saturation regime, employing the  QP or FP profiles can be advantageous. For the QP profile, by changing the focal parameter, the optimal position of the IRS can be adjusted. For the FP profile, the IRS can achieve the optimal outage performance for a large range of Tx-to-IRS distances rather than at a single point.   
\end{itemize}  
To the best of the authors' knowledge,  a power scaling law for optical IRSs with LP profile and a performance comparison  with  relays  were first reported in \cite{Globecom_ScalingLaw}, which is the conference version of this paper. In contrast to \cite{Globecom_ScalingLaw}, in this paper, we do not only consider   IRSs with  LP profile but also with QP  and FP   profiles  and investigate the impact of the phase shift profile on the power scaling law. Moreover, the  optimal positions of  IRSs with QP and FP profiles were not studied in \cite{Globecom_ScalingLaw}. 

\textit{Notations:} Boldface lower-case and upper-case letters denote vectors and matrices, respectively.  Superscript $(\cdot)^T$ and $\mathbb{E}\{\cdot\}$
denote the transpose and expectation operators, respectively. $x\sim \mathcal{N}(\mu,\sigma^2)$ represents a  Gaussian random variable with mean $\mu$ and variance $\sigma^2$. $\mathbf{I}_n$ is the $n\times n$ identity matrix, $j$ denotes the imaginary unit, and $(\cdot)^*$ and $\mathcal{R}\{\cdot\}$ represent the complex conjugate and real part of a complex number, respectively. Moreover, $\text{erf}(\cdot)$ and $\text{erfi}(\cdot)$ are the error function and the imaginary error function, respectively. Here, $\Phi(x)=\int_{-\infty}^{x}\frac{1}{\sqrt{2\pi}} e^{-\frac{t^2}{2}} \mathrm{d}t$ denotes the cumulative distribution function (CDF) of the univariate normal density. Furthermore, $\mathbb{R}^+$ denotes the set of positive real numbers. $\lVert\cdot\rVert$ denotes the vector $\ell_2$-norm and $\text{sinc}(x)=\frac{\sin(x)}{x}$ is the sinc function.
\vspace*{-2mm}
\section{System and Channel Models}\label{Sec_System}

	\begin{figure}[t!]
		\centering
			\includegraphics[width=0.65\textwidth]{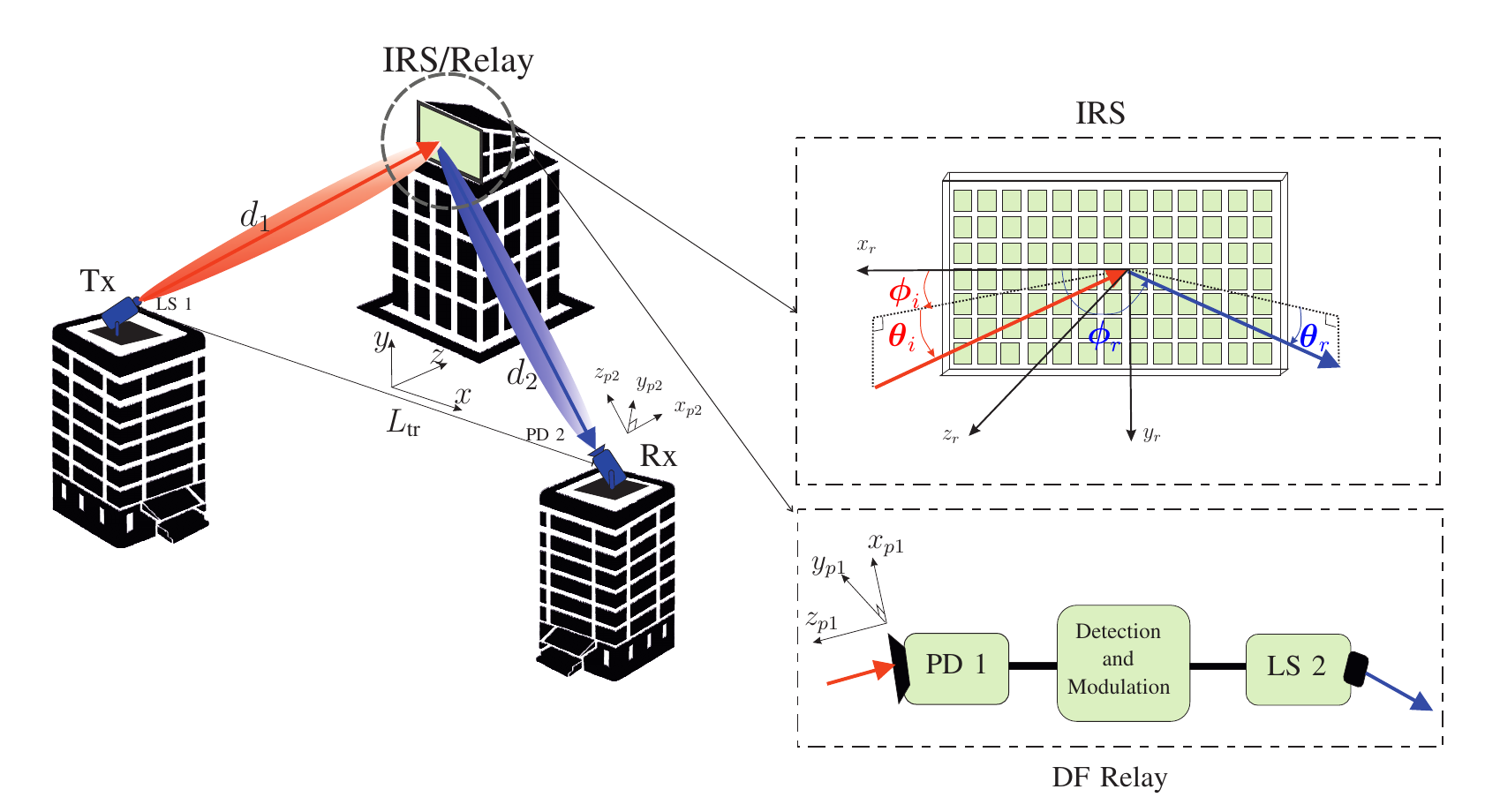}
		\caption{IRS- and relay-based FSO systems.}
		\label{Fig:System}\vspace{-0.8cm}
	\end{figure} 

We consider two FSO systems, where Tx and  Rx are connected via an IRS and a relay, respectively.  Tx and  Rx are located  on the $x$-axis with  distance  $L_\text{tr}\over 2$ from the origin of the $xyz$-coordinate system, respectively, see Fig.~\ref{Fig:System}. Moreover, the centers of the IRS and the relay are located at the origin of the $x_ry_rz_r$-coordinate system, where the $x_ry_r$-plane  is parallel to the $xy$-plane and the $z_r$-axis points in the opposite direction  of the $z$-axis, see Fig.~\ref{Fig:System}.  The Tx is equipped with laser source (LS) 1 emitting a Gaussian laser beam. The beam axis  intersects with the $x_ry_r$-plane at distance  $d_1$ and points in direction $\mathbf{\Psi}_{i}=\left(\theta_{i},\phi_{i}\right)$, where $\theta_i$ is the angle between the $x_ry_r$-plane and the beam axis, and $\phi_{i}$ is the angle between the projection of the beam axis on the $x_ry_r$-plane and the  $x_r$-axis. Moreover, the Rx is equipped with photo-detector (PD) 2 and  a circular  lens of radius $a$.  The lens  of the Rx   is located at distance $d_{2}$ from the origin of the $x_ry_rz_r$-coordinate system. The normal vector of  the lens plane points in direction $\mathbf{\Psi}_{r}=\left(\theta_{r}, \phi_{r}\right)$, where $\theta_{r}$ is the angle between the $x_ry_r$-plane and the normal vector, and $\phi_{r}$ is the angle between the projection of the normal vector on the $x_ry_r$-plane and the $x_r$-axis. Furthermore, we assume that Tx, Rx,  IRS, and relay  are installed at the same height, i.e., $\phi_{i}=0$ and $\phi_{r}=\pi$, and the  Rx lens plane is always perpendicular to the  axis of the received beam.  In the following, we describe the  relay- and IRS-assisted FSO systems more in detail.
\subsection{Relay-Assisted FSO System Model}
We assume  the Tx is connected to the Rx via a full-duplex DF relay\footnote{In this work, we assume  DF relaying, as under typical channel conditions, it achieves  better or at least  similar performance  compared to AF relaying \cite{Sahar_Placement}.}, where the relay receives the transmitted symbols via PD 1, re-encodes the signal, and transmit it via LS 2   to  PD 2  at the Rx.  PD 1  is equipped with a circular lens of radius $a$, which is always perpendicular to the axis of the received beam. 
Assuming an intensity modulation and direct detection (IM/DD) system,  the received signal   at the relay, $y_{1}$, is given by 
\begin{IEEEeqnarray}{rll}
y_{1}&=\sqrt{P_1}h_1 s_1+n_{1},
	\label{system_eq_SR}
\end{IEEEeqnarray}
where $s_1$ is  the  symbol transmitted by LS 1   with $\mathbb{E}\{|s_1|^2\}=1$, $P_1$ is the transmit power of LS 1, $h_1\in \mathbb{R}^+$  denotes the  gain of the Tx-relay link,  and $n_1\sim \mathcal{N} \left(0, \sigma_n^2\right)$ is the additive white Gaussian noise (AWGN) at PD 1 with zero mean and variance $\sigma_n^2$. Then, the received signal   at  PD 2, $y_{2}$, is given by
\begin{IEEEeqnarray}{rll}
y_{2}&=\sqrt{P_2}h_2 s_2+n_2,
	\label{system_eq_RD}
\end{IEEEeqnarray}
where  $s_2$ is the  signal transmitted by the relay with $\mathbb{E}\{|s_2|^2\}=1$, $h_{2}\in \mathcal{R}^+$ denotes the FSO channel gain of the relay-Rx link,  $P_2$ is the transmit power of LS 2, and $n_2\sim \mathcal{N} \left(0, \sigma_n^2\right)$ is the AWGN noise at PD 2. For simplicity, $P_1$ and $P_2$ are chosen such that   $P_1=P_2=\frac{P_\text{tot}}{2}$, where $P_\text{tot}$ is the total transmit power. 
\subsection{IRS-Assisted FSO System Model}\label{Sec_PhaseProfiles}
We assume the Tx  is connected via a single IRS to the Rx. The size of the IRS is $\Sigma_\text{irs}=L_{x}\times L_{y}$,   where $L_x$ and $L_y$ are the lengths of the IRS in  $x_r$- and $y_r$-direction, respectively. 
Optical IRSs can be implemented using mirror-based or MM-based technologies \cite{Vahid_Magazine}. Here, we consider the following setups:
\begin{itemize}[leftmargin=*]
	\item \textbf{Mirror-based IRSs:} A single conventional mirror or an array of small micro-mirrors can be employed to  provide specular reflection, i.e., the incident and reflected beam angles w.r.t.  the IRS are identical.  To provide the desired reflection angle, the mirror can be re-oriented with a mechanical motor, which rotates the mirror  with elevation and azimuth  angles $\Theta_\text{mir}=\frac{|\theta_r-\theta_i|}{2}$ and $\varphi_\text{mir}=\frac{|\pi-\phi_r+\phi_i|}{2}$, respectively.  Since $\phi_{i}=0$ and $\phi_r=\pi$, the azimuth angle is given by $\varphi_\text{mir}=0$.
	\item \textbf{MM-based IRSs:} An MM-based IRS is a planar surface which is comprised of  passive sub-wavelength elements to manipulate the properties of the incident beam. Since typically $L_x,L_y\gg  \lambda$, where $\lambda$ is the wavelength, MM-based IRSs can be modeled as  continuous surfaces with  continuous  phase shift profile. The following phase shift designs are considered in this paper:
\setlist[enumerate]{leftmargin=*}
\begin{enumerate}
\item \textit{Linear phase shift (LP) profile:} Anomalous reflection and redirection of the  beam originating from the LS towards the lens is accomplished with an IRS  phase shift profile  which changes linearly along the $x_r$- and $y_r$-axes as follows \cite{IRS_FSO_WCNC}:
 \begin{IEEEeqnarray}{rll}
	\Phi_\text{irs}({\mathbf{r}_r})=k\left(\Phi_x x_r+\Phi_y y_r+\Phi_0\right),
	\label{EQ:Phase_LP}
\end{IEEEeqnarray} 
where $k=\frac{2\pi}{\lambda}$ is the wave number and $\mathbf{r}_r=(x_r,y_r,0)$ denotes a point in the $x_ry_r$-plane. To redirect the beam from  Tx  direction $\mathbf{\Psi}_i$ to Rx direction $\mathbf{\Psi}_r$, the phase shift gradients are chosen as  $\Phi_x=\cos(\theta_{i})\cos(\phi_{i})+\cos(\theta_r)\cos(\phi_r)$, $\Phi_y=\cos(\theta_{i})\sin(\phi_i)+\cos(\theta_r)\sin(\phi_r)$, and  $\Phi_0$ is constant.
\item \textit{Quadratic phase shift  (QP) profile:} Focusing the beam at distance $f$ from the IRS along  direction $\mathbf{\Psi}_r$ and reducing the beamwidth of the reflected beam can be realized with  a phase shift profile which changes quadratically along the $x_r$- and $y_r$-axes, respectively. Thus, we consider the following phase shift profile which eliminates the accumulated phase of the incident beam and adds a parabolic phase shift \cite{TCOM_IRSFSO}:
\begin{IEEEeqnarray}{rll}
	\Phi_\text{irs}(\mathbf{r}_r)
	=k\left(\Phi_{x^2}x_r^2+\Phi_{y^2}y_r^2+\Phi_{x}x_r+\Phi_{y}y_r+\Phi_0\right),
	\label{EQ:Phase_QP}
\end{IEEEeqnarray}
where    $\Phi_{x^2}=-\frac{\sin^2(\theta_{i})}{2R(d_1)}-\frac{\sin^2(\theta_{r})}{2d_2}+\frac{\sin^2(\theta_{r})}{4f}$ and $\Phi_{y^2}=-\frac{1}{2R(d_1)}-\frac{1}{2d_2}+\frac{1}{4f}$. 
\item \textit{Focusing phase shift (FP) profile:} An ideal phase shift profile, which focuses the  incident  beam power on the lens center, is given as follows
\begin{IEEEeqnarray}{rll}
	\Phi_\text{irs}(\mathbf{r}_r)=-\psi_\text{in}-k\lVert \tilde{\mathbf{r}}_o-\mathbf{r}_r \rVert,
		\label{EQ:Phase_FP}
\end{IEEEeqnarray}
where  $\tilde{\mathbf{r}}_o=(\tilde{x}_o,\tilde{y}_o,\tilde{z}_o)=(-d_2\cos\theta_p, 0, d_2 \sin\theta_p)$ denotes the center of the lens and $\psi_\text{in}$ is the phase of the incident  beam on the IRS.
\end{enumerate}
\end{itemize} 
The above considered technologies and phase shift profiles  enhance the end-to-end FSO link performance as follows:   An MM-based IRS with LP profile behaves similarly to a mirror in redirecting the beam towards the Rx but the latter can direct more power towards the Rx lens because of the larger effective IRS aperture achieved by its reorientation facilitated by mechanical motors. Moreover, an IRS with QP profile can  reduce the beamwidth at the lens and  mitigate  beam divergence along the propagation path. Furthermore, an IRS with  FP profile  yields a very narrow beam  at the lens such that the lens can collect most of the transmitted power in the absence of misalignment and tracking errors.   

Assuming an IM/DD FSO system, regardless of the adopted phase shift profile,   for an IRS-assisted FSO link the  received signal  $y_{3}$ at  PD 2, is given by
\begin{IEEEeqnarray}{rll}
	y_{3}=\sqrt{P_\text{tot}}h_3 s_1 +n_2,
	\label{system_eq_IRS}
\end{IEEEeqnarray}
where   $h_3 \in \mathcal{R}^+$ is the end-to-end channel gain between Tx, IRS, and Rx  and  LS 1 transmits with  power $P_\text{tot}$.
\subsection{Channel Model}
FSO channels are impaired by  geometric and misalignment losses (GML),  atmospheric loss, and  atmospheric turbulence induced fading \cite{IRS_FSO_WCNC}.   Thus, the point-to-point FSO channel gains of the considered systems can be modeled as follows
\begin{IEEEeqnarray}{rll}
	h_i=\zeta h_{p,i} h_{a,i} h_{\text{gml},i}, \quad i\in\{1,2,3\},
	\label{ch}
\end{IEEEeqnarray}
where $\zeta$ is the PD responsivity, $h_{a,i}$ represents the random atmospheric turbulence induced fading,  $h_{p,i}$ is the atmospheric loss, and  $h_{\text{gml},i}$ characterizes the GML. 
\subsubsection{Atmospheric Loss}
The atmospheric loss characterizes the laser beam energy loss due to absorption and scattering and  is given by	$h_{p,i}=10^{-\frac{\kappa d_i}{10}},  i\in\{1,2,3\},$ where $\kappa$ is the attenuation coefficient and   $d_{3}=d_1+d_2$ denotes the end-to-end link distance.
\subsubsection{Atmospheric Turbulence}
The variations of the refractive index along the propagation path due to changes in temperature and pressure cause atmospheric turbulence which is analogous to  the fading in RF systems. Assuming $h_{a,i}$ is  a Gamma-Gamma distributed random variable, its cumulative distribution function  (CDF)  is given by \cite{Alouini-GG}
\begin{IEEEeqnarray}{rll}
	F_{h_{a,i}}(x)=\frac{1}{\Gamma(\alpha_i)\Gamma(\beta_i)} G_{1,3}^{2,1}\left(\alpha_i\beta_i{{x}}\left|\begin{array}{@{}l@{}} \ \ \ \ \ \ \ 1 \\ \alpha_i,\beta_i,0 \\ \end{array} \right. \right),
	\label{EQ:CDF_GG}
\end{IEEEeqnarray}
where $\Gamma(\cdot)$ denotes the Gamma function and  $G_{p,q}^{m,n}\left(\varkappa\left|\begin{array}{@{}l@{}}  e_1,\dots e_p \\f_1, \dots, f_q \\ \end{array} \right.\right)$ is the Meijer G-function  \cite{integral}.  Here,  the small and large scale turbulence parameters $\alpha_i$ and $\beta_i$  depend on the Rytov variance $\sigma_{R,i}^2=1.23C_n^2k^{7\over 6}d_i^{11\over 6}$, where  $C_n^2$ is the refractive-index structure constant  \cite{Sahar_Placement}.  Given that $d_3>d_1,d_2$, the Ryotov variance of the IRS-assisted link is larger, which leads to more severe fading compared to the relay-assisted link.

\subsubsection{GML}
The GML coefficient comprises the deterministic geometric loss due to the divergence of the laser beam along the transmission path and the random misalignment loss due to transceiver sway \cite{Vahid_Magazine}. Here,  we neglect the misalignment  loss 
\footnote{In  practice, the misalignment loss  can be avoided or considerably reduced  by using sophisticated  acquisition and tracking mechanisms such as gimbals, mirrors, and adaptive optics \cite{ATP_Ansari}.} 
and determine the geometric loss for the relay- and IRS-based links in the following.

Assuming the waist of the Gaussian beam is larger than the wavelength,  the electric field of the Gaussian laser beam  emitted by the  $i$-th LS, $\forall i\in\{1, 2\}$,  is given by \cite{Goodmanbook}
\begin{IEEEeqnarray}{rll}
E_{\ell i}\!\left(\mathbf{r}_{\ell i}\right)&= C_\ell e^{-{x_{\ell i }^2+y_{\ell i}^2\over w^2(z_{\ell i })}-j	k\left(\!\!z_{\ell i}+{x_{\ell i}^2+y_{\ell i }^2\over 2R(z_{\ell i })}\right)-\tan^{-1}\left(\frac{z_{\ell i}}{z_{R i}}\right)},\qquad
	\label{Gauss}
\end{IEEEeqnarray}
where $C_\ell=\sqrt{\frac{4\eta P_\text{tot}}{n\pi{w}^2(z_{\ell i})}}$, $\eta$ is the free-space impedance, $n=1$ and  $2$ for the IRS- and relay-assisted links, respectively, and $\mathbf{r}_{\ell i}=(x_{\ell i }, y_{\ell i }, z_{\ell i})$ is a point in a  coordinate system, which has its origin at the $i$-th LS.  The $z_{\ell i }$-axis of this coordinate system  is along the beam axis, the $y_{\ell i }$-axis is parallel to the intersection line of the $i$-th LS plane and the $x_ry_r$-plane, and the $x_{\ell i}$-axis  is orthogonal to the $y_{\ell i}$- and $z_{\ell i}$-axes.   $w(z_{\ell i})=w_{o i}\Big[{1+\left(\frac{z_{\ell i}}{z_{R i}}\right)^2}\Big]^{1/2}$ is the beamwidth at distance $z_{\ell i}$, $w_{oi}$ is the beam waist,  $z_{R i}=\frac{\pi w_{o i}^2}{\lambda}$ is the Rayleigh range, and $R(z_{\ell i})=z_{\ell i}\Big[1+\left(\frac{z_{R i}}{z_{\ell i}}\right)^2\Big]$ is the radius of the curvature of the beam's wavefront.

Assuming the lenses at the relay and the Rx are always perpendicular to the incident beam axes, respectively, the GML coefficients  of the  Tx-relay link, $h_{\text{gml},1}$, and the relay-Rx link, $h_{\text{gml},2}$, are given by \cite{Farid_fso}
\begin{IEEEeqnarray}{rll}
	h_{\text{gml},i}={1 \over  \eta P_\text{tot}}\iint\nolimits_{\mathcal{A}_{pi}}\lvert E_{\ell i}\left(\mathbf{r}_{pi}\right) \rvert^2 \, \mathrm{d}\mathcal{A}_{pi},  \,\quad i\in\{1,2\},
	\label{gain_rel}
\end{IEEEeqnarray}
 where $\mathcal{A}_{pi}$  denotes the area of the lens of  PD $i$ and $\mathbf{r}_{pi}=(x_{pi}, y_{pi}, z_{pi})$ denotes a   point in the $i$-th lens plane.  The origin of the $x_{pi}y_{pi}z_{pi}$-coordinate system is  the center of the  $i$-th lens and the $z_{pi}$-axis  is parallel to  the normal vector of the $i$-th lens plane. We assume that the $y_{pi}$-axis is parallel to the intersection line of the lens plane and the $x_ry_r$-plane and the $x_{pi}$-axis is perpendicular to the $y_{pi}$- and $z_{pi}$-axes.
Moreover, the GML coefficient of the IRS-assisted FSO link is given by \cite{TCOM_IRSFSO}
\begin{IEEEeqnarray}{rll}
	h_{\text{gml},3}={1 \over 2 \eta P_\text{tot}}\iint\nolimits_{\mathcal{A}_{p2}}\lvert E_{r}\left(\mathbf{r}_{p2}\right) \rvert^2 \, \mathrm{d}\mathcal{A}_{p2}, \quad
	\label{gain}
\end{IEEEeqnarray}
where $E_{r}\left(\mathbf{r}_{p2}\right)$ is  the electric field of the  beam reflected by the IRS and received at the Rx lens  and  is given by  \cite{TCOM_IRSFSO}
 
\begin{IEEEeqnarray}{rll}
	E_{r}\left(\mathbf{r}_{p2}\right)=C_r\iint_{({x}_r,{y}_r)\in\Sigma_\text{irs}} E_{\text{in}}({\mathbf{r}_r}) \exp\left(-jk \lVert\mathbf{r}_{o}-\mathbf{r}_r\rVert \right) e^{-j\Phi_\text{irs}(\mathbf{r}_r)} \mathrm{d}{x}_r\mathrm{d}{y}_r. \quad
	\label{Huygens}
\end{IEEEeqnarray}
Here, $C_r={\sqrt{\sin(\theta_r)}}/({j\lambda d_2})$, $\mathbf{r}_o=\left(\mathbf{r}_{p2}+\mathbf{c}_1\right)\mathbf{R}_\text{rot}$,  $\mathbf{R}_\text{rot}=\left(\begin{smallmatrix}
-\sin(\theta_r) &0 &-\cos(\theta_r)\\
0&-1&0\\
-\cos(\theta_r)&0&\sin(\theta_r)
\end{smallmatrix}\right)$, $\mathbf{c}_1=(0,0,d_2)$, and $E_{\text{in}}({\mathbf{r}_r})$ is the   electric field incident on the IRS given in \cite[Eq.~(7)]{TCOM_IRSFSO} as follows
	\begin{IEEEeqnarray}{rll}
	E_\text{in}(\mathbf{r}_r)&=C_\ell \sqrt{|\sin(\theta_i)|}e^{-\frac{x_r^2}{w^2_{\text{in},x}}-\frac{y_r^2}{w^2_{\text{in},y}}-j\psi_\text{in}(\mathbf{r}_r)} \quad\text{with phase}\nonumber\\
	\psi_\text{in}(\mathbf{r}_r)&=k\left(d_1-x_r\cos(\theta_{i})+\frac{x_r^2\sin^2(\theta_{i})}{2R(d_1)}+\frac{y_r^2}{2R(d_1)}\right)-\tan^{-1}\left(\frac{d_1}{z_{R1}}\right),
	\label{EQ:E_incident}
\end{IEEEeqnarray}
where $w_{\text{in},x}=\frac{w(d_1)}{\sin(\theta_i)}$ and $w_{\text{in},y}=w(d_1)$ are the incident beamwidths in the IRS plane in $x$- and $y$-direction, respectively.  A closed-form solution of (\ref{gain}) for IRSs with LP and QP profiles operating at intermediate- and far-field distances (Fresnel regime), i.e., $d_2\gg d_n$, where $d_n=\left[\frac{(x_e^2+y_e^2)(x_e+y_e)}{4\lambda}\right]^{\frac{1}{2}}$ and $i_e = \min(\frac{L_i}{2}, w_{\text{in},i})$, $i\in\{x,y\}$, is given by \cite[Eq.~(20)]{TCOM_IRSFSO}\footnote{For a typical FSO link with $w_{o1}=2.5$ mm, $d_1=d_2=700$ m, $\theta_{i}=\theta_{r}=\frac{\pi}{4}$ and square-shaped IRS with length $0.5$ m, we have $x_e=0.2$ m, $y_e=0.14$ m, which leads to  $d_n=55$ m. This means that for distances $d_2>55\, \text{m}$, the  result in \cite[Eq.~(20)]{TCOM_IRSFSO}} is valid.   
  Unfortunately, \cite[Eq.~(20)]{TCOM_IRSFSO} is a very involved expression and cannot be used to derive the dependence  of the received power on the  IRS and lens sizes and also does not provide insight into the    end-to-end system performance. Thus, in the following, we reanalyze (\ref{gain})  for different IRS  and lens sizes to derive power scaling law for  optical IRS.
\section{ Power Scaling Law for  Optical IRSs }\label{Sec:Power Regimes}
In this section, we analyze the received power at the Rx lens for mirror-based and MM-based IRSs with different phase shift profiles. The received power at the lens for the IRS-assisted link using (\ref{system_eq_IRS}) and  (\ref{ch}) is given by
\begin{IEEEeqnarray}{rll}
	P_\text{rx}^\text{irs}=P_\text{tot}\zeta^2 h_{p,3}^2h_{a,3}^2 h_{\text{gml},3}^2,
\end{IEEEeqnarray}
where only the GML factor $h_{\text{gml},3}$   depends on the IRS and lens sizes. 
In this paper, to gain insight for FSO system design and to determine the corresponding power scaling law, we analyze $h_{\text{gml},3}$ given by (\ref{gain}) and (\ref{Huygens}) for different IRS sizes, $\Sigma_\text{irs}$, and lens sizes, $\Sigma_\text{lens}$, as follows:
\begin{itemize} 
	\item Regime 1: $\Sigma_\text{irs}\ll A_\text{in}$ and $\Sigma_\text{lens}\ll A_\text{rx}^\text{g1}$
	\item Regime 2: $\Sigma_\text{irs}\ll A_\text{in}$ and $\Sigma_\text{lens}\gg A_\text{rx}^\text{g2}$
	\item Regime 3: $\Sigma_\text{irs}\gg A_\text{in}$
\end{itemize}
where $A_\text{in}=\pi w_{\text{in},x}w_{\text{in},y}$ and $A_\text{rx}^{q}=\pi w_{\text{rx},x}^{\iota,q}w_{\text{rx},y}^{\iota,q}$ are the areas of the equivalent beam footprints in the IRS and Rx lens planes, respectively. Moreover,  $w_{\text{rx},x}^{\iota,q}$ and $w_{\text{rx},y}^{\iota,q}$ are the  equivalent received beam widths in the lens plane in $x_p$- and $y_p$-direction, respectively. Here, ${q}\in\{\text{g1}, \text{g2},\text{g3}\}$ denotes the operating regime of the IRS and $\iota\in\{\text{LP}, \text{QP}, \text{FP}, \text{mir}\}$ indicates the  IRS technology and phase shift profile.  In the following, we show  that depending on the system parameters,  the GML, and hence, the received power  scales quadratically or linearly with the IRS size or it may remain constant.
\subsection{Regime 1: Quadratic Power Scaling Regime}\label{Sec:Power Regimes_Lin}
In this regime, we  consider the case where the IRS is small such that  only a fraction of the Gaussian incident beam is captured by the IRS and the beam can be approximated by a plane wave. We derive the GML coefficient using the plane wave approximation  in Lemma \ref{Lemma1}. Then, for a lens smaller than the received beam footprint, the GML coefficient is approximated in Corollary \ref{Corollary1}. 
\begin{lem}\label{Lemma1}
	Assuming an Rx lens at distances $d_2\gg d_n$  and an IRS with
	\begin{itemize}
		\item LP profile and   $L_x\ll \min\{ w_{\text{in},x},\sqrt{\frac{2d_1d_2}{kd_3}}\}$, $L_y\ll \min\{w_{\text{in},y},\sqrt{\frac{2d_1d_2}{kd_3}}\}$
		\item QP profile and   $L_x\ll \min\{w_{\text{in},x},\sqrt{\frac{4f}{k}}\}$ and $L_y\ll \min\{ w_{\text{in},y},\sqrt{\frac{4f}{k}}\}$
		\item FP profile and   $L_x\ll w_{\text{in},x}$ and $L_y\ll w_{\text{in},y}$
	\end{itemize}
		the  GML coefficient for the IRS-assisted  link, $h_{\text{gml},3}$,  can be approximated by $\tilde{G}_1^\iota$, which is given as follows	
\begin{IEEEeqnarray}{rll}
	 		\tilde{G}_1^\iota=C_1&\left[\frac{a \sqrt{\pi}}{w_{\text{rx},x}^{\iota,\text{g1}}} \text{Si}\left(\frac{a \sqrt{\pi}}{w_{\text{rx},x}^{\iota,\text{g1}}}\right)+\cos\left(\frac{a \sqrt{\pi}}{w_{\text{rx},x}^{\iota,\text{g1}}} \right)-1\right]\nonumber\\
	 		\times&\left[\frac{a \sqrt{\pi}}{w_{\text{rx},y}^{\iota,\text{g1}}} \text{Si}\left(\frac{a \sqrt{\pi}}{w_{\text{rx},y}^{\iota,\text{g1}}}\right)+\cos\left(\frac{a \sqrt{\pi}}{w_{\text{rx},y}^{\iota,\text{g1}}}\right)-1\right],\quad \iota\in\{\text{LP},\text{QP},\text{FP}\},
	 		\label{EQ:Lem1}
\end{IEEEeqnarray}
	where $C_1=\frac{8 d_2^2\lambda^2\left|\sin(\theta_i)\right|}{\pi^6 a^2  w^2(d_1) \left|\sin(\theta_r)\right|}$,  $\text{Si}(\varkappa)=\int_{0}^{\varkappa}\frac{\sin(t)}{t} \mathrm{d}t$ denotes the sine integral function, and   $w_{\text{rx},x}^{\iota,\text{g1}}=\frac{2d_2}{k\sin(\theta_r)L_x}$, $w_{\text{rx},y}^{\iota,\text{g1}}=\frac{2d_2}{kL_y}$ can be interpreted as equivalent beamwidths.
\end{lem}  
\begin{IEEEproof}
	The proof is given in Appendix \ref{App1}.
\end{IEEEproof}
Lemma \ref{Lemma1} reveals that due to the small IRS size, the GML coefficient is independent of the considered IRS phase shift profile. Furthermore, due to the small IRS size, the amplitude of the received electric field is the product of two sinc-functions, see (\ref{EQ:Proof_Lem_Quad_LP_I}) in Appendix \ref{App1}.  Thus,  the coherent superposition of the signals reflected from all points on the IRS at the lens introduces a beamforming gain.  By increasing the IRS size,   the beamwidth of the sinc-shaped beam at the lens decreases, which in turn  increases the peak amplitude of the beam causing the beamforming gain. In addition to this beamforming gain, a larger IRS surface collects more power from the incident beam which results in a quadratic  scaling of the power received at the Rx lens with the IRS size. This behavior is analytically confirmed in the following corollary. 
\begin{corol}\label{Corollary1}
	For $\Sigma_\text{lens}\ll A_\text{rx}^\text{g1}$ and $d_1\gg z_{R1}$, $\tilde{G}_1^\iota$ can be approximated by
	\begin{IEEEeqnarray}{rll}
	{G}_1^\iota&= \frac{16\pi^2 \Sigma_\text{irs}^2 \left|\sin(\theta_r)\right|\left|\sin(\theta_i)\right| }{\lambda^4} \times g_{\text{LS}} \times g_\text{PD}, \quad \iota\in\{\text{LP},\text{QP},\text{FP}\},
		\label{EQ:ApproxG_1}
	\end{IEEEeqnarray}
where $g_\text{LS}=\frac{2\pi w_{o1}^2}{4\pi d_1^2}$ and $g_\text{PD}=\frac{\pi a^2}{4\pi d_2^2}$. Since $G_1^\iota$ scales quadratically with the  IRS size, $\Sigma_\text{irs}$, we refer to this regime as the ``quadratic power scaling regime''.
\end{corol}
\begin{IEEEproof}
	 Assuming $a\sqrt{\pi}\ll \min\{ w_{\text{rx},x}^{\iota,\text{g1}},w_{\text{rx},y}^{\iota,\text{g1}}\}$, we  substitute in (\ref{EQ:Lem1}) the Taylor series expansions of $\text{Si}(x)\approx x$ and $\cos(x)\approx 1-\frac{x^2}{2}$. Then, assuming $d_1\gg z_{R1}$, we can substitute  $w(d_1)\approx \frac{d_1 \lambda}{\pi w_{o1}}$. This completes the proof.
\end{IEEEproof}  
The quadratic scaling law shown in Corollary \ref{Corollary1} is in agreement with the power scaling law  in  \cite[Eq.~(2), (10)]{Marzieh_Poor} and \cite[Eq.~(48)]{Emil-PowerScale} for RF IRSs. 

\begin{corol}\label{Col:LP_mir}
	Assuming a  mirror-based IRS of size $\Sigma_\text{irs}\ll \min\{A_\text{in},\frac{2d_1d_2}{kd_3}\}$, which can mechanically rotate around the $z_r$-axis with rotation angle $\Theta_\text{mir}=\frac{|\theta_r-\theta_i|}{2}$, and an Rx lens size  $\Sigma_\text{lens}\ll A_\text{rx}^\text{g1}$, the  GML for a mirror-assisted  link, $h_{\text{gml},3}$,  can be approximated by ${G}_1^\text{mir}$, which is given as follows
	\begin{IEEEeqnarray}{rll}
		{G}_1^\text{mir}&= \frac{16\pi^2 \Sigma_\text{irs}^2 \left|\sin(\theta_i^\text{mir})\right|^2 }{\lambda^4} \times g_{\text{LS}} \times g_\text{PD}. \quad
		\label{EQ:ApproxG_1_mir}
	\end{IEEEeqnarray}
	where $\theta_i^\text{mir}=\frac{\theta_i+\theta_r}{2}$. 
\end{corol}
 \begin{IEEEproof}
	Given the rotation angle of the mirror  $\Theta_\text{mir}$, we can substitute $\theta_i$ and $\theta_{r}$ in (\ref{Huygens}) with $\theta_{i}^\text{mir}$ and $\theta_r^\text{mir}$, respectively, where $\theta_i^\text{mir}=\theta_r^\text{mir}$ due to the specular reflection. Then, for a mirror,  $\Phi_\text{irs}=0$   and following similar steps as in Appendix \ref{App1}  leads to (\ref{EQ:ApproxG_1_mir}). This completes the proof. 
\end{IEEEproof}
\subsection{ Regime 2: Linear Power Scaling Regime}\label{Sec:G2}
As the size of the IRS increases, eventually  the beamforming gain of the IRS saturates and cannot further increase the received power.   In this regime, the lens is much larger than the beam footprint at the Rx such that the total  power incident on the IRS is received at the Rx lens. Thus, by further increasing the IRS size,  the received power  increases only linearly. In the following lemma, we provide the GML  for this case.

\begin{lem}\label{Lemma2}
	Assuming  $a\gg \min\{w_{\text{rx},x}^{\iota,\text{g2}}, w_{\text{rx},y}^{\iota,\text{g2}}\}$, where $w_{\text{rx},i}^{\iota,\text{g2}}=\frac{2\sqrt{\tilde{b}_{i,\iota}}}{k}, \forall i\{x,y\}$, are the equivalent beamwidths in the linear power regime, 	the  GML factor for an IRS with phase shift profile $\iota$, $h_{\text{gml},3}^\iota$,  can be approximated by $\tilde{G}_2^\iota$, which is given as follows
\begin{IEEEeqnarray}{rll}
	\tilde{G}_2^\iota=&\frac{1}{16}\left[8\text{T}(a_{1},c_{2,1})+8\text{T}(a_{1}^*,c_{2,1}^*)+4\right]\left[8\text{T}(a_2,c_{2,2})+8\text{T}(a_2^*,c_{2,2}^*)+4\right], \iota\in\{\text{LP},\text{QP},\text{FP}\},\quad
	\label{EQ:Lem2}
\end{IEEEeqnarray} 
where $\text{T}(\cdot,\cdot)$ is the Owen's T function  \cite{Owen_Integ_book}, $a_{m}=\frac{\sqrt{2}\zeta_{1,m}}{\sqrt{1+2\zeta_{2,m}^2}}$,  $c_{2,m}=-\frac{\zeta_{1,m}^*\left(1+2\zeta_{2,m}^2+2\frac{\zeta_{1,m}}{\zeta_1^*}|\zeta_{2,m}|^2\right)}{\zeta_{1,m}\sqrt{1+2\zeta_{2,m}^2+2(\zeta_{2,m}^*)^2}}$, $m\in\{1,2\}$,  $\zeta_{1,1}=\frac{\sqrt{b_{x,\iota}}L_x}{2}$, $\zeta_{2,1}=j\frac{\sqrt{b_{x,\iota}^*}}{2\sqrt{B_{x,\iota}}}$,${\zeta}_{1,2}=\frac{\sqrt{b_{y,\iota}}L_y}{2}$,  ${\zeta}_{2,2}=j\frac{\sqrt{b_{y,\iota}^*}}{2\sqrt{B_{y,\iota}}}$, $b_{x,\text{LP}}=\frac{\sin^2(\theta_i)}{w^2(d_1)}+\frac{jk\sin^2(\theta_i)}{2R(d_1)}+\frac{jk\sin^2(\theta_r)}{2d_2}$, $b_{y,\text{LP}}=\frac{1}{w^2(d_1)}+\frac{jk}{2R(d_1)}+\frac{jk}{2d_2}$, $b_{x,\text{QP}}=\frac{\sin^2(\theta_i)}{w^2(d_1)}+\frac{jk\sin^2(\theta_r)}{4f}$,  $b_{y,\text{QP}}=\frac{1}{w^2(d_1)}+\frac{jk}{4f}$, $b_{x,\text{FP}}=\frac{\sin^2(\theta_{i})}{w^2(d_1)}$,  $b_{y,\text{FP}}=\frac{1}{w^2(d_1)}$, $B_{i,\iota}=\mathcal{R}\{b_{i,\iota}\}$, and 
$\tilde{b}_{i,\iota}=\frac{{b}_{i,\iota}{b}_{i,\iota}^*}{{b}_{i,\iota}+{b}_{i,\iota}^*}$.
\end{lem} 
\begin{IEEEproof}
	The proof is provided in Appendix \ref{App2}.
\end{IEEEproof}
In (\ref{EQ:Lem2}), by assuming large lens sizes, $\tilde{G}_2^\iota$   only depends on the IRS size and phase shift profile. To unveil the second power scaling regime, we approximate (\ref{EQ:Lem2}) in the following corollary.
\begin{corol}
	Assuming  $\Sigma_\text{irs}\ll A_\text{in}$,    $\tilde{G}_2^\iota$ can be approximated by
	\begin{IEEEeqnarray}{rll}
		\bar{G}_2^{\iota}&=\text{erf}\left(\frac{\sqrt{2}}{2}\frac{L_x\sin(\theta_i)}{w(d_1)}\right)\text{erf}\left(\frac{\sqrt{2}}{2}\frac{L_y}{w(d_1)}\right).\quad
		\label{EQ:ApproxG_22}
	\end{IEEEeqnarray}	
\end{corol}
\begin{IEEEproof}
	We approximate $c_{2,i}\to -\infty$ and we exploit $\text{T}(x,\infty)=\frac{1-\Phi(|x|)}{2}$ in \cite[pp. 414, Eq.~(2.4)]{Owen_Integ_book} to obtain  $\tilde{G}_2^\iota=\frac{1}{16}\left[-4+8\Phi(|a_1|)\right]\left[-4+8\Phi(|a_2|)\right]$.
	Substituting $\Phi(x)=\frac{1}{2}\text{erf}\left(\frac{x}{\sqrt{2}}+1\right)$, we obtain (\ref{EQ:ApproxG_22}) and this completes the proof.
\end{IEEEproof}

\begin{corol}\label{Col:Linear}
	Assuming  $\frac{L_x}{w_{\text{in},x}}, \frac{L_y}{w_{\text{in},y}}\to 0$,    $\bar{G}_2^\iota$ can be further approximated by
	\begin{IEEEeqnarray}{rll}
		{G}_2^{\iota}&= \frac{4\pi \Sigma_\text{irs} \left|\sin(\theta_i)\right|}{\lambda^2}\times g_\text{LS}.\quad
		\label{EQ:ApproxG_2}
	\end{IEEEeqnarray}	
	Since $G_2^\iota$ scales linearly with the IRS size, $\Sigma_\text{irs}$,  we refer to this regime as the ``linear power scaling regime''.
\end{corol}
\begin{IEEEproof}
	We  apply the Taylor series expansion of $\lim\limits_{x\to 0}\text{erf}(x)=\frac{2}{\sqrt{\pi}}x$ in (\ref{EQ:ApproxG_22}) to obtain (\ref{EQ:ApproxG_2}). This completes the proof.	
\end{IEEEproof}
As can be observed from Corollary \ref{Col:Linear}, similar to the quadratic power scaling regime, in this regime,  the collected power at the Rx lens does  also not depend on the phase shift profile of the IRS.
\begin{corol}
Assuming a mirror-based IRS and  IRS and lens sizes of  $\Sigma_\text{irs}\ll A_\text{in}$ and $\Sigma_\text{lens}\gg A_\text{rx}^\text{g2}$,  respectively, the GML factor is given by
\begin{IEEEeqnarray}{rll}
	{G}_2^\text{mir}=  \frac{4\pi \Sigma_\text{irs} \left|\sin(\theta_i^\text{mir})\right|}{\lambda^2}\times g_\text{LS}.\quad
	\label{EQ:ApproxG_2_mir}
\end{IEEEeqnarray}
\end{corol}
\begin{IEEEproof}
	Following the same steps as in Corollary \ref{Col:LP_mir} and \ref{Col:Linear} leads to (\ref{EQ:ApproxG_2_mir}), which completes the proof.
\end{IEEEproof}
\subsection{Regime 3: Saturated Power Scaling Regime}
For the case, when the IRS size is very large, such that the lens size is the limiting factor for the received power,  the GML for an IRS with LP profile  is given in  the following lemma.
\begin{lem}\label{Lemma3}
	Assuming  $\Sigma_\text{irs}\gg A_\text{in}$ and the LP profile in (\ref{EQ:Phase_LP}), the GML coefficient, $h_{\text{gml},3}$,  can be approximated by $G_3^\text{LP}$, which is given by
	\begin{IEEEeqnarray}{rll}
		G_3^\text{LP}=\text{erf}\left(\sqrt{\frac{\pi}{2}}\frac{a}{w_{\text{rx},x}^{\text{LP}, \text{g3}}}\right)\text{erf}\left(\sqrt{\frac{\pi}{2}}\frac{a}{w_{\text{rx},y}^\text{LP}}\right),
		\label{EQ:Lem3}
	\end{IEEEeqnarray}
 where\footnote{We notice that the equivalent received beamwidths for IRSs in the  linear and saturation regimes are identical, i.e.,  $w_{\text{rx},i}^{\iota, \text{g3}}=w_{\text{rx},i}^{\iota, \text{g2}}$. This is because we have ignored the impact of diffraction for calculating  the equivalent beamwidths.} $w_{\text{rx},x}^{\text{LP}, \text{g3}}=\frac{w(d_1)\left|\sin(\theta_r)\right|}{\left|\sin(\theta_i)\right|} \left[\left(\frac{\Lambda_1 \sin^2(\theta_{i})}{ \sin^2(\theta_{r})}\right)^2+\left(\frac{\Lambda_2 \sin^2(\theta_{i})}{ \sin^2(\theta_{r})}+1\right)^2\right]^{1/2}\!\!\!\!\!\!$, $w_{\text{rx},y}^{\text{LP}, \text{g3}}=w(d_1) \left[\Lambda_1^2+\left(\Lambda_2 +1\right)^2\right]^{1/2}\!\!\!\!$, $\Lambda_1=\frac{2d_2}{kw^2(d_1)}$, and $\Lambda_2=\frac{d_2}{R(d_1)}$.
\end{lem} 
\begin{IEEEproof}
	The proof is provided in Appendix \ref{App3}.
\end{IEEEproof}
The above lemma shows that, in the considered case, the normalized received power at the lens does not depend on the IRS size. Thus, we refer to this regime as the ``saturation power scaling regime''. Moreover, $\Lambda_1$ reflects the increase of the beamwidth along the propagation path due to diffraction  and $\Lambda_2$ reflects the refraction effect and the divergence  of the reflected beam, which further increases the received beamwidth at the lens. We note that for the plane waves typical for far-field  RF links  $\Lambda_1=\Lambda_2=0$ would hold.
\begin{corol}
	Assuming $d_1 \gg z_{R1}$  and that the center of the IRS with LP profile  is positioned at distance $d_1=d_2=\frac{d_3}{2}$ and direction $\theta_{i}=\theta_{r}$, then, the GML coefficient  is given by
	\begin{IEEEeqnarray}{rll}
		G_3^\text{LP}=\left[\text{erf}\left(\sqrt{\frac{\pi}{2}}\frac{a}{w(d_{3})}\right)\right]^2.
		\label{EQ:Lem3_Approx}
	\end{IEEEeqnarray}	
\end{corol}
\begin{IEEEproof}
 Substitute $d_1=d_2=\frac{d_{3}}{2}$ and $\theta_{i}=\theta_r$ in (\ref{EQ:Lem3}). Thus, using $d_{3} \gg z_{R1}$, we can approximate $w(d_1)\approx \frac{\lambda d_1}{\pi w_{o1}}$ and $R(d_1)\approx d_1$. Thus, $w_{\text{rx},x}^{\text{LP}, \text{g3}}=w_{\text{rx},y}^{\text{LP}, \text{g3}}=w_{o1}\sqrt{1+\left(\frac{d_{3}}{z_{R1}}\right)^2}= w(d_{3})$, which leads to (\ref{EQ:Lem3_Approx}). This completes the proof.
\end{IEEEproof}
The above corollary shows that when the IRS has identical distance to  LS and  lens, the incident and reflection angles are identical, and if the  IRS  is larger than the beam illuminating  its surface, then, the received beamwidth at the lens is $w(d_3)$. Thus, in this case, the end-to-end IRS-assisted link behaves similar to a point-to-point  FSO link of length $d_3$.

Next, for an IRS with QP profile,  given the large IRS size in the considered regime, the IRS   can collect more power   by increasing the focal distance and decreasing the received beamwidth as shown in the following Lemma.
\begin{lem}\label{Lemma_sat_QP}
	Assuming   $\Sigma_\text{irs}\gg A_\text{in}$ and the QP profile in (\ref{EQ:Phase_QP}), the GML coefficient, $h_{\text{gml},3}^\text{QP}$,  can be approximated by 
	\begin{IEEEeqnarray}{rll}
		G_3^\text{QP}=\text{erf}\left(\sqrt{\frac{\pi}{2}}\frac{a}{w_{\text{rx},x}^{\text{QP}}}\right)\text{erf}\left(\sqrt{\frac{\pi}{2}}\frac{a}{w_{\text{rx},y}^{\text{QP}}}\right),
		\label{EQ:G3_QP}
	\end{IEEEeqnarray}
	where the equivalent beamwidths for the QP profile are given by $w_{\text{rx},y}^{\text{QP}, \text{g3}}=w(d_1)\sqrt{\Lambda_1^2+\left(\frac{d_2}{2f}\right)^2}$ and $w_{\text{rx},x}^{\text{QP}, \text{g3}}=w(d_1)\frac{\sin(\theta_{r})}{\sin(\theta_{i})}$$\sqrt{\left(\frac{\sin^2(\theta_{i})}{\sin^2(\theta_{r})}\Lambda_1\right)^2+\left(\frac{d_2}{2f}\right)^2}$.
\end{lem} 
\begin{IEEEproof}
	The proof is provided in Appendix \ref{App8}.
\end{IEEEproof}
As can be observed from Lemma \ref{Lemma_sat_QP}, by increasing the focal distance $f$, the beam footprint in the lens plane gets smaller. Thus,  by adjusting $f$, the beamwidth on the lens can be optimized to improve  performance. Moreover, by comparing (\ref{EQ:G3_QP}) and (\ref{EQ:Lem3}), we can show that  by choosing $f=\frac{kw^2(d_1) \sin^2(\theta_r)}{2\sqrt{2}\sin(\theta_{i})\sqrt{\sin^2(\theta_i)-2\sin^2(\theta_{r})}}$, an IRS with QP profile  behaves identically to an IRS with LP profile.

\begin{lem}\label{Lemma_sat_FP}
	Assuming $\Sigma_\text{irs}\gg A_\text{in}$ and the FP profile in (\ref{EQ:Phase_FP}), the GML coefficient, $h_{\text{gml},3}^\text{FP}$,  can be approximated by $G_3^\text{FP}$, which is given by
	\begin{IEEEeqnarray}{rll}
		G_3^\text{FP}= \text{erf}\left(\sqrt{\frac{\pi}{2}}\frac{a}{w_{\text{rx},x}^{\text{FP}, \text{g3}}}\right) \text{erf}\left(\sqrt\frac{{\pi }}{2} \frac{a}{w_{\text{rx},y}^{\text{FP}, \text{g3}}}\right),
		\label{EQ:Lem_sat_FP}	
	\end{IEEEeqnarray}
	where $w_{\text{rx},x}^{\text{FP}, \text{g3}}=\frac{{2}d_2 |\sin(\theta_{i})|}{kw(d_1)|\sin(\theta_r)|}$ and $w_{\text{rx},y}^{\text{FP}, \text{g3}}=\frac{{2}d_2 }{kw(d_1)}$.
\end{lem}
\begin{IEEEproof}
	The proof is provided in Appendix \ref{App11}.
\end{IEEEproof}
As can be observed from (\ref{EQ:Lem_sat_FP}), by substituting $w(d_1)\approx \frac{\lambda d_1}{\pi w_{o1}}$ in $w_{\text{rx},i}^\text{FP}, \,i\in\{x,y\}$,   the received beamwidth at the lens is on the order of $w_{o1}$ which is much smaller than the lens radius $a$. This leads to $G_3^\text{FP}\approx 1$, which is the maximum possible GML coefficient for an IRS-assisted FSO link. Moreover, by comparing  (\ref{EQ:G3_QP}) and (\ref{EQ:Lem_sat_FP}), we conclude that an IRS with  QP profile  with   $f=\infty$   performs identical to an IRS with FP profile. 
\begin{corol}
 Assuming $\Sigma_\text{irs}\gg A_\text{in}$, the GML coefficient of a mirror-assisted IRS, $h_{\text{gml},3}$, is given by
\begin{IEEEeqnarray}{rll}
	G_3^\text{mir}=\left[\text{erf}\left(\sqrt{\frac{\pi}{2}}\frac{a}{w_{\text{rx}}^{\text{mir}, \text{g3}}}\right)\right]^2,
	\label{EQ:Lem3_mir}
\end{IEEEeqnarray}
where $w_{\text{rx}}^{\text{mir}, \text{g3}}=w(d_1)\left[\Lambda_1^2+\left(\Lambda_2+1\right)^2\right]^{\frac{1}{2}}$.
\end{corol}
\begin{IEEEproof}
Following the same steps as in Corollary \ref{Col:LP_mir} and Lemma \ref{Lemma3} leads to (\ref{EQ:Lem3_mir}) which completes the proof.
\end{IEEEproof}

\subsection{GML Coefficients}
Here, we summarize the GML coefficients for IRS- and relay-assisted FSO links.
\subsubsection{ GML Coefficient of IRS-Assisted FSO Link ($h_{\text{gml},3}$)}
To determine the boundary  IRS sizes at which the power scaling regime changes, we shall  analytically derive the received beamwidth at the lens\footnote{The exact beamwidth at the lens can be obtained as $w_{\text{rx},i}^\text{exact}=\frac{W_{F,i}}{\sqrt{2\ln(2)}}$. Here, $W_F$ is the beamwidth, where the power intensity is half the maximum intensity. Let us consider the IRS with FP profile in (\ref{EQ:I_FP}). To obtain $W_{F,i}$, we shall find $\mathbf{r}_p$ such that  $I_r(\mathbf{r}_p)=\frac{1}{2}\left|\text{erf}\left(\frac{L_x}{2w_{\text{in},x}}\right)\text{erf}\left(\frac{L_y}{2w_{\text{in},y}}\right)\right|$. Because of the $\text{erf}$-terms, it is not easy to obtain a closed-form solution for $W_{F,i}$.}. In order to simplify the calculations, we propose approximate boundary values  for which the quadratic, linear, and saturation power scaling laws are valid.  
\begin{prop}\label{Theorem1}
	If  $G_3^\iota\geq \frac{2d_2^2w_{o1}^2|\sin(\theta_i)|}{d_1^2a^2|\sin(\theta_r)|}$,  $h_{\text{gml},3}^i$  scales   with the IRS size,  $\Sigma_\text{irs}$,  as follows
	\begin{IEEEeqnarray}{rll}
		h_{\text{gml},3}^\iota\approx
		\begin{cases}
			{G}_1^\iota, \quad &  \Sigma_\text{irs}< S_1,\\
			{G}_2^\iota, \quad &  S_1\leq \Sigma_\text{irs}\leq S_2^\iota, \\
			G_3^\iota, \quad & \Sigma_\text{irs} > S_2^\iota,
		\end{cases}		\qquad \iota\in\{\text{LP},\text{QP}, \text{FP}, \text{mir}\},
		\label{EQ:Theo1}
	\end{IEEEeqnarray}
where $S_1=\frac{\lambda^2 d_2^2}{\pi a^2\left|\sin(\theta_r)\right|}$ and $S_2^\iota= \frac{\pi G_3^\iota w^2(d_1)}{2\left|\sin(\theta_{i})\right|}$ are the boundary IRS sizes, where the  transition from  quadratic to  linear and from  linear to saturation power scaling  occurs, respectively. If 
	$S_2^\iota<S_1$, i.e., $G_3^\iota< \frac{2d_2^2w_{o1}^2|\sin(\theta_i)|}{d_1^2a^2|\sin(\theta_r)|}$, the GML scales only quadratically  with the IRS size,  $\Sigma_\text{irs}$, or is  constant, i.e.,
	\begin{IEEEeqnarray}{rll}
		h_{\text{gml},3}^\iota\approx
		\begin{cases}
			{G}_1^\iota ,\quad & \Sigma_\text{irs}\leq S_3^\iota,\\
			G_3^\iota , \quad & \Sigma_\text{irs}> S_3^\iota,
		\end{cases}		\qquad \iota\in\{\text{LP},\text{QP}, \text{FP}, \text{mir}\},
		\label{EQ:Theo1b}
	\end{IEEEeqnarray}
	where   $S_3^\iota=\frac{\sqrt{G_3^\iota}\lambda d_2 w(d_1)}{a \sqrt{2\sin(\theta_i)\sin(\theta_r)}}$ is the IRS size for which the transition from  quadratic to linear power scaling  occurs.
\end{prop} 
\begin{IEEEproof}
	Boundary $S_1$ is derived  as the intersection point of (\ref{EQ:ApproxG_1}) and (\ref{EQ:ApproxG_2}) and similarly  $S_2^\iota$  is the intersection point of  $G_3^\iota$ and (\ref{EQ:ApproxG_2}). If $S_2^\iota<S_1$,  linear power scaling does not occur and thus, $S_3^\iota$ is the intersection point of (\ref{EQ:ApproxG_1})  and $G_3^\iota$. This leads to (\ref{EQ:Theo1b}) and completes the proof.
\end{IEEEproof}
The above proposition  shows  how the received power scales with the IRS size for given system parameters such as the LS parameters (i.e., $w_{o1}$ and $\lambda$),  lens radius (i.e., $a$),    distances (i.e., $d_1$ and $d_2$), and  the incident and reflected angles (i.e., $\theta_i$ and $\theta_r$). Moreover, due to the large electrical size of the Rx lens ($\frac{\pi a^2}{\lambda^2}\approx 10^{8}$), the boundary IRS size, $S_1$, is comparatively small, and thus, optical IRSs of sizes $10$ cm-1 m   typically operate  in the linear  or saturation power scaling regimes.  Unlike FSO systems, the electrical size of  RF receive antennas is comparatively small (e.g., dipole $\approx 1$, antenna arrays $\approx 100$) which leads to large values for $S_1$, and thus, even RF IRSs with large sizes of $1-10$ m   operate in the quadratic power scaling regime \cite{Emil-PowerScale}.
\begin{figure}[t]
	\centering
		\includegraphics[width=0.6\textwidth]{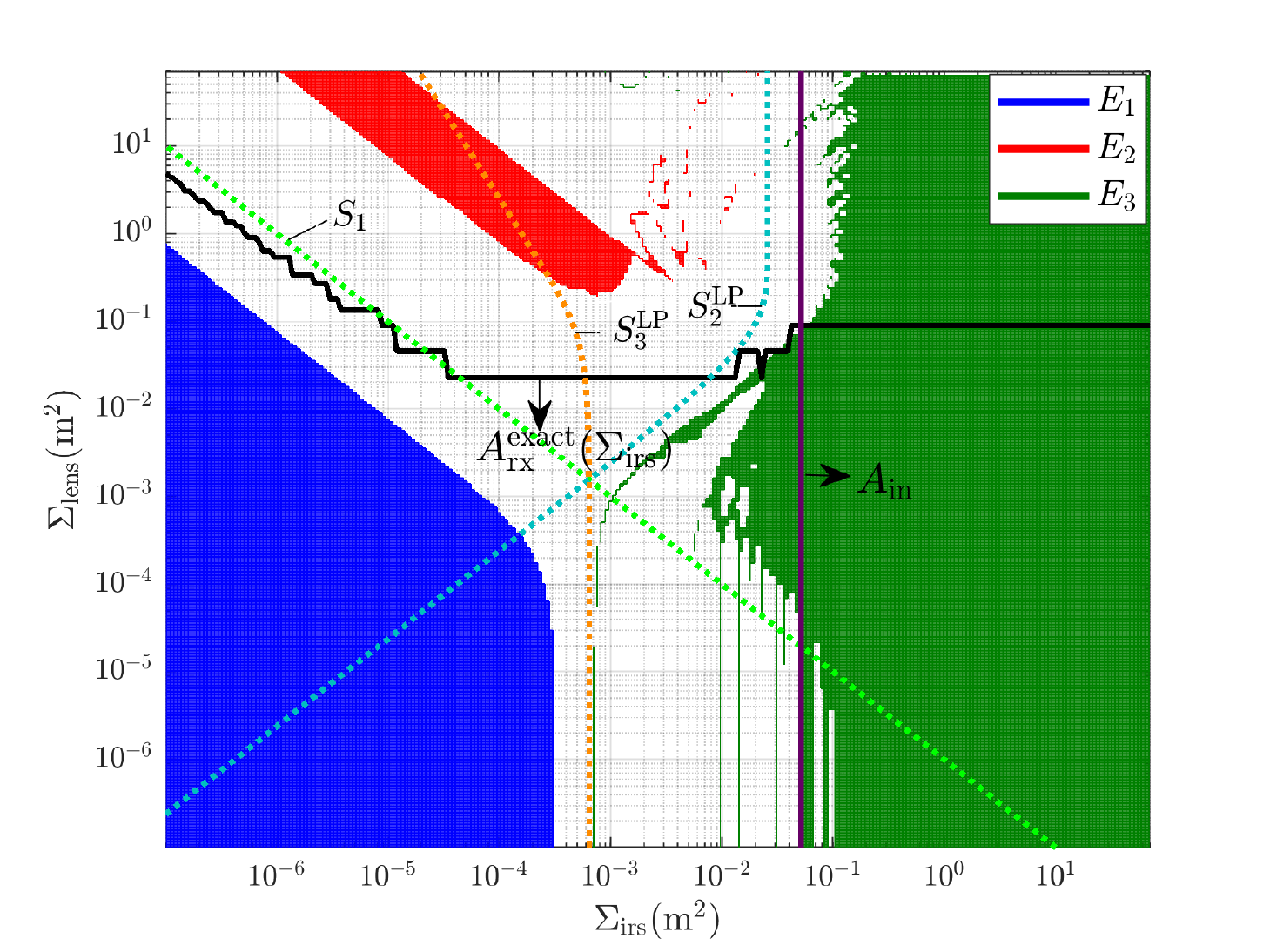}
	\vspace{-4mm}
	\caption{Normalized error, $E_i=\frac{|G_i^\text{LP}-h_{\text{gml},3}^\text{LP}|}{\max(G_i^\text{LP},h_{\text{gml},3}^\text{LP})},\, i\in{1,2,3}$, between the simulated GML, $h_{\text{gml},3}^\text{LP}$,  and the analytical GML, $G_i^\text{LP}$. The black line shows the exact beam footprint area on the lens, denoted by $A_\text{rx}^\text{exact}(\Sigma_\text{irs})=\pi w_{\text{rx},x}^{\text{exact}}w_{\text{rx},y}^\text{exact}$,  versus $\Sigma_\text{irs}$. Regions with  normalized error values $E_i\leq 0.05$ are shown in red, blue, and green.\label{Fig:GML_3D}} 
\end{figure}

For IRSs with LP profile, Fig.~\ref{Fig:GML_3D} shows the regions where the normalized error  between $h_{\text{gml},3}^\text{LP}$ obtained by (\ref{gain}), (\ref{Huygens}) and the analytical GML coefficients in (\ref{EQ:Theo1}) is less than $0.05$.  The three proposed power scaling regimes  are evident in  this figure: 1) Quadratic regime (blue region): The GML factor $G_1^\text{LP}$ in (\ref{EQ:ApproxG_1}) matches  $h_{\text{gml},3}^\text{LP}$ in the region where $\Sigma_\text{irs}\ll A_\text{in}$ and $\Sigma_\text{irs}\ll A_\text{rx}(\Sigma_\text{irs})$. 2) Linear regime (red region): The GML factor $G_2^\text{LP}$ in (\ref{EQ:ApproxG_2}) matches  $h_{\text{gml},3}^\text{LP}$ in the region where $\Sigma_\text{irs}\ll A_\text{in}$ and $\Sigma_\text{irs}\gg A_\text{rx}(\Sigma_\text{irs})$. 3) Saturation regime (green region): The GML factor $G_3^\text{LP}$ in (\ref{EQ:Lem3}) matches  $h_{\text{gml},3}^\text{LP}$ in the region where $\Sigma_\text{irs}\gg A_\text{in}$. 
Curves $S_1$, $S_2^\text{LP}$, and $S_3^\text{LP}$ indicate the boundary IRS sizes versus the lens size. As can be observed, for lens sizes $\Sigma_\text{lens}\geq16\, \mathrm{cm}^2$, curves $S_1$ and $S_2^\text{LP}$ define the boundary IRS sizes between the three power scaling regimes as given in (\ref{EQ:Theo1}). However,  for $\Sigma_\text{lens}<16\, \mathrm{cm}^2$,   curve $S_3$ defines the boundary between quadratic and saturation regimes as given in (\ref{EQ:Theo1b})\footnote{We note that the boundaries of the colored regimes in Fig.~\ref{Fig:GML_3D} move closer to $S_1$, $S_2^\iota$, and $S_3^\iota$ if larger error values are allowed, i.e., for $E_i>0.05$. }.

\subsubsection{GML in Relay-Based FSO Link}
For  relay-based FSO links,  we assume that the lenses at the relay and the Rx are always  orthogonal to the axis of the respective incident beam. Thus, by substituting the electric field of the LS in (\ref{Gauss})  and  solving the integral in (\ref{gain_rel}), we obtain 
\begin{IEEEeqnarray}{rll}
	h_{\text{gml},i}=\left[\text{erf}\left(\sqrt{\frac{{\pi}}{2}}\frac{a}{w(d_i)}\right)\right]^2, \quad i\in\{1,2\},
	\label{EQ:hgrelay}
\end{IEEEeqnarray}
for the two involved links, which matches the deterministic GML of  point-to-point FSO links in \cite{Farid_fso}. 
\section{Comparison of IRS- and Relay-based Links}
In this section, we compare the outage probability performance of IRS- and relay-assisted links and derive  the diversity and coding gains. Next, we determine the optimal positions of the IRS and relay to minimize the outage probability at high SNR values. 

\subsection{Diversity and Coding Gains}\label{Sec_Diversity}
For a fixed transmission rate, the outage probability is defined as the probability that the instantaneous SNR, $\gamma$,   is smaller than a threshold SNR, $\gamma_{th}$, i.e., $P_\text{out}=\text{Pr}\left(\gamma<\gamma_{th}\right)$. At high SNR, the outage probability can be approximated as $\lim\limits_{\bar{\gamma}\to\infty}P_\text{out}\approx(C\bar{\gamma})^{-D}$, where $C$ is the coding gain,  $\bar{\gamma}$ is the average transmit SNR, and $D$ is the diversity gain. In the following, we compare the  diversity and coding gains of IRS- and relay-assisted FSO systems. 
\subsubsection{Outage Performance of IRS-assisted Link}
For the IRS-assisted FSO link in (\ref{system_eq_IRS}), the average received power is $\bar{\gamma}_3=\bar{\gamma} \tilde{\gamma}_3$, where $\bar{\gamma}=\frac{P_\text{tot}}{\sigma_n^2}$ and $\tilde{\gamma}_3=h_{\text{gml},3}^2h_{p,3}^2$, and thus, the outage probability is given by \cite{Sahar_Placement}
\begin{IEEEeqnarray}{rll}
	P_\text{out}^\text{irs}=F_{h_{a,3}}\left(\sqrt{{\gamma_{th}}/{\bar{\gamma}_3}}\right),
	\label{EQ:Outage_IRS}
\end{IEEEeqnarray}
where $F_{h_{a,3}}(\cdot)$ is given in (\ref{EQ:CDF_GG}). Thus, using the same approach as in \cite{Sahar_Placement}, the diversity gain, $D_\text{irs}$, and the coding gain, $C_\text{irs}$, of an IRS-assisted FSO link  respectively can be obtained as
\begin{IEEEeqnarray}{rll}
D_\text{irs}=\frac{\varrho_3}{2},\quad
C_\text{irs}=\frac{\tilde{\gamma}_3}{\gamma_{th}(\tau_3 \varrho_3)^2}\left(\frac{\Gamma\left(\tau_3-\varrho_3\right)}{\Gamma(\tau_3)\Gamma(\varrho_3+1)}\right)^{-1/D_\text{irs}},
\label{EQ:Diversity_IRS}	
\end{IEEEeqnarray}	
where $\varrho_3=\min\{\alpha_3, \beta_3\} $ and $\tau_3=\max\{\alpha_3, \beta_3\}$. 
\subsubsection{Outage Performance of Relay-assisted Link}
The outage probability of a relay-assisted FSO link is given by \cite{Sahar_Placement}
\begin{IEEEeqnarray}{rll}
	P_\text{out}^\text{rel}=1-\prod_{i=1}^{2}\left(1-F_{h_{a,i}}\left(\sqrt{{\gamma_{th}}/{\bar{\gamma}_i}}\right)\right),
	\label{EQ:Outage_relay}
\end{IEEEeqnarray}
where $\bar{\gamma}_i=\bar{\gamma}\tilde{\gamma}_i$ and $\tilde{\gamma}_i=\frac{1}{2}h_{\text{gml},i}^2h_{p,i}^2$, $\forall i\in\{1,2\}$.
 Moreover,  the  diversity gain, $D_\text{rel}$, and the coding gain, $C_\text{rel}$,  of a relay-assisted FSO link are given as follows \cite{Sahar_Placement}
\begin{IEEEeqnarray}{rll}
	D_\text{rel}&=\min\{\frac{\varrho_1}{2},\frac{\varrho_2}{2}\},\quad\, C_\text{rel}&=\begin{cases}
		C_{\text{rel},\upsilon}&\!\!\!\varrho_1\neq\varrho_2,\\
		\!\!\left(\!\sum\limits_{i=1}^{2}\left(C_{\text{rel},i}\right)^{-D_\text{rel}}\!\!\right)^{\!\!\!-1\over D_\text{rel}}&\!\!\!\varrho_1=\varrho_2,	
	\end{cases}\!\!, 
	\label{EQ:Diversity_Relay}	
\end{IEEEeqnarray}	
 where $C_{\text{rel},i}=\frac{\tilde{\gamma}_i}{\gamma_{th}}\left(\frac{\Gamma(\tau_i-\varrho_i)\left(\tau_i\varrho_i/\mu_i\right)^{\varrho_i}}{\Gamma(\tau_i)\Gamma(\varrho_i+1)}\right)^{-2/\varrho_i}$,  $\varrho_i=\min\{\alpha_i, \beta_i\} $, $\tau_i=\max\{\alpha_i, \beta_i\}$, $\forall i\in\{1,2\}$, and $\upsilon=\mathrm{arg} \min\limits_{i\in\{1,2\}}\{\varrho_i\}$. 

For larger distances, the Gamma-Gamma  fading parameters, $\alpha_i$ and $\beta_i$, become smaller, see \cite[Eq.~(33)]{Sahar_Placement}. Thus, $\varrho_3<\min\{\varrho_1,\varrho_2\}$, and thus, according to (\ref{EQ:Diversity_IRS}) and (\ref{EQ:Diversity_Relay}), the diversity gain of a relay-assisted link is larger than that of an IRS-assisted link. Thus, a relay-assisted link   outperforms an IRS-assisted  link at high SNRs. However,  depending on the system parameters, the coding gain of the IRS-assisted FSO link  may be larger than that of the relay-assisted link, which can boost the performance at low SNRs, see numerical results in Section \ref{Sec_Sim}.
 
\subsection{Optimal Operating Position of IRS and Relay}\label{Sec:Placement}
	\begin{figure}[t!]
	\centering
			\includegraphics[width=0.6\textwidth]{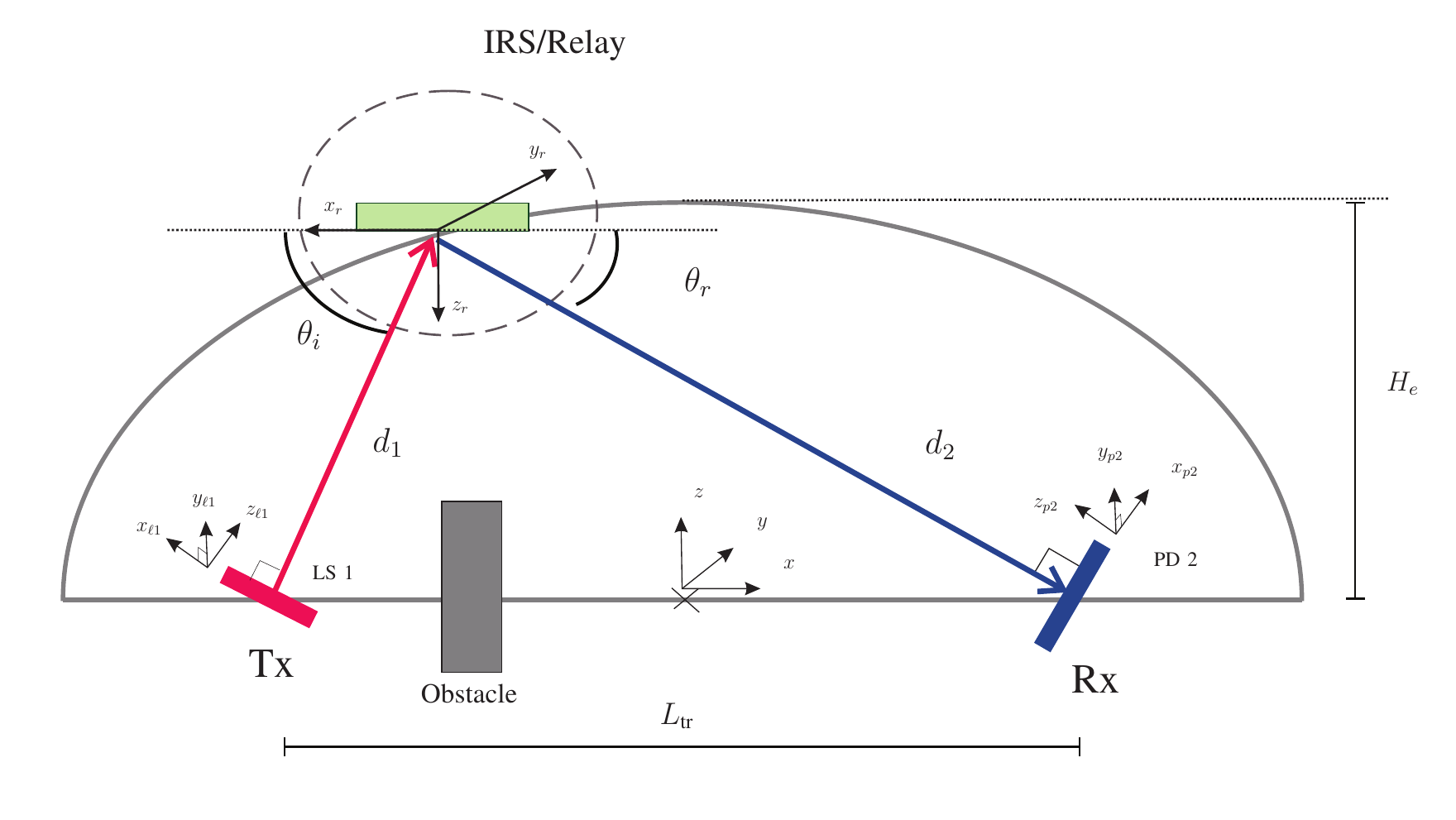}
	\caption{IRS- and relay-based FSO systems on an ellipse with width $d_3$ and height $H_e$.}
	\label{Fig:elp}\vspace{-0.8cm}
\end{figure} 
Exploiting the analysis in Sections \ref{Sec:Power Regimes} and \ref{Sec_Diversity}, we determine the optimal  positions of the center of the IRS and relay, denoted by $(x_o^*, z_o^*)$ in the $xyz$-coordinate system, where the outage probabilities of the IRS- and relay-assisted links  at high SNR are minimized, respectively. For a fair comparison, we assume that the end-to-end distance $d_3$ is constant, i.e., IRS and relay are located on an ellipse, see Fig.~\ref{Fig:elp}. Thus, we formulate the following optimization problem
\begin{IEEEeqnarray}{rll}
	&\min_{x_o, z_o} P_\text{out}^i\approx \left(C_i \bar{\gamma}\right)^{-D_i},\quad i\in\{\text{rel}, \text{irs}\},\nonumber\\
	&\text{s.t.}\quad \frac{x_o^2}{d_3^2}+\frac{z_o^2}{d_3^2-L_\text{tr}^2}=\frac{1}{4}.
	\label{EQ:Opt_prob}
\end{IEEEeqnarray}
\subsubsection{Optimal Position of  IRS}
Given that   parameters $\alpha_3$ and $\beta_3$   only depend on the end-to-end distance \cite{IRS-FSO-Herald, TCOM_IRSFSO},  the outage probability of the IRS-assisted link in (\ref{EQ:Opt_prob}) is minimized if  $h_{\text{gml},3} $ is maximized. The optimal position of the IRS as a function of its size is given in the following theorem.

\begin{theo}\label{Lemma5}
	The  optimal position of the center of the IRS, $(x_o^*, z_o^*)$,  depends on the size  of the IRS and the phase shift profile,  and is given by 
	\begin{IEEEeqnarray}{rll}
		(x_o^*,z_o^*)\!\!=\!\!
		\begin{cases}
		 \left(\pm\frac{\sqrt{2\rho_1}d_{3}}{4L_\text{tr}},z^*_1 \right), &\Sigma_\text{irs}\leq \min \{S_1, S_3^\iota\}, \\
		 \left(\frac{d_{3}}{8L_\text{tr}}\left(d_{3}- \rho_2\right), z^*_2 \right), 	&S_1\leq \Sigma_\text{irs}\leq S_2^\iota,\\
		 \left(x_o(d_1^\iota), z_o(d_1^\iota)\right),&\max\{S_2^\iota,S_3^\iota\} \leq \Sigma_\text{irs},
		\end{cases}	\quad \iota\in\{\text{LP},\text{QP},\text{FP}, \text{mir}\}
	\label{EQ:theo2}
	\end{IEEEeqnarray}
where $\rho_1=3L_\text{tr}^2-d_{3}^2$, $\rho_2=\sqrt{d_3^2+24L_\text{tr}^2}$, $z^*_1=H_e[1-\frac{\rho_1}{L_\text{tr}^2}]^{1\over2}$,  $z^*_2=H_e\left[1-\frac{1}{8L_\text{tr}^2}\left(d_{3}^2+12L_\text{tr}^2-\rho_2d_{3}\right)\right]$, and $H_e=\frac{1}{2}\sqrt{d_3^2-L_\text{tr}^2}$. Moreover, $d_1^\text{LP}=\frac{d_3}{2}$ and  $d_1^\text{QP}$ is the solution to the following  equation
 \begin{IEEEeqnarray}{rll}
&\frac{\text{erf}\left(\frac{\sqrt{\pi}}{2}\frac{a}{\omega_1(d_1)}\right)}{\text{erf}\left(\frac{\sqrt{\pi}}{2}\frac{a}{\omega_2(d_1)}\right)}\frac{\omega_1^4(d_1)}{\omega_2^4(d_1)}e^{-\frac{\pi a^2}{2}\left(\frac{1}{\omega_1^2(d_1)}-\frac{1}{\omega_2^2(d_1)}\right)} \left(-4d_3z_{R1}^2 (d_3-d_1)^3f^2+d_1^8\right)\nonumber\\
&-2z_{R1}^2d_1^2f^2(d_3-d_1)d_3 +\frac{1}{2}d_1^6(d_3-d_1)(-2d_1+d_3)=0,
\label{EQ:Optimal_pos_eq_QP}
\end{IEEEeqnarray}
where $\omega_1(d_1)=\frac{\lambda}{\pi w_{o1}} \sqrt{z_{R1}^2\frac{(d_3-d_1)^4}{d_1^4}+\frac{d_1^4}{4f^2}}$ and $\omega_2(d_1)=\frac{\lambda}{\pi w_{o1}} \sqrt{z_{R1}^2\frac{(d_3-d_1)^2}{d_1^2}+\frac{(d_3-d_1)^2d_1^2}{4f^2}}$.
Moreover, $d_1^\text{FP}$ is the solution to the following  equation
 \begin{IEEEeqnarray}{rll}
2\varkappa	e^{-\vartheta^2 \varkappa^4}\text{erf}(\vartheta\varkappa)+
 e^{-\vartheta^2 \varkappa^2}\text{erf}(\vartheta\varkappa^2)=0,
	\label{EQ:Optimal_pos_eq_FP}
\end{IEEEeqnarray}
where $\varkappa=\frac{d_1}{d_3-d_1}$ and $\vartheta=\sqrt{\frac{\pi}{2}}\frac{a}{w_{o1}}$.

\end{theo}
\begin{IEEEproof}
	The proof is provided in Appendix \ref{App5}.
\end{IEEEproof}
The above theorem suggests that for small  IRSs  operating in the quadratic power scaling regime, the optimal position of the IRS is  close to  Tx or Rx,  which is in agreement with the results for RF IRSs in \cite{RF-OptPos, Marzieh_Poor}. Moreover,  IRSs operating in the linear power scaling regime achieve better performance close to the Tx. However, when the IRS size is  large, the optimal position of  the IRS 
depends on its phase shift profile. For large IRSs with LP profile, the optimal position is  equidistant from  Tx and Rx. For the QP profile, the optimal position can be controlled by using parameter $f$ in (\ref{EQ:Optimal_pos_eq_QP}). For the FP profile, the optimal position is given by (\ref{EQ:Optimal_pos_eq_FP}). Since in practice the lens radius is always larger than the beam waist, i.e.,  $a\gg w_{o1}$, then, for $\kappa>1$ (or equivalently $d_1>\frac{d_3}{2}$), (\ref{EQ:Optimal_pos_eq_FP}) is always valid. Thus,  an IRS with FP profile operating in the saturation regime achieves optimal outage performance for a range of Tx-to-IRS distances,  $d_1>\frac{d_3}{2}$, rather than at a single optimal position. 
\begin{theo}\label{Lemma6}
	The  optimal position of the center of the mirror, $(x_o^*, z_o^*)$,  depends on the size  of the mirror   and is given by 
	\begin{IEEEeqnarray}{rll}
		(x_o^*,z_o^*)\!\!=\!\!
		\begin{cases}
			\left(\pm\frac{d_3}{2},0 \right), &\Sigma_\text{irs}\leq \min \{S_1, S_3^\text{mir}\}, \\
			\left(\frac{d_{3}}{2},0 \right), 	&S_1\leq \Sigma_\text{irs}\leq S_2^\text{mir},\\
		\left([-\frac{d_3}{2},\frac{d_3}{2}],[0,H_e]\right),&\max\{S_2^\text{mir},S_3^\text{mir}\} \leq \Sigma_\text{irs},
		\end{cases}	\quad 
		\label{EQ:theo3}
	\end{IEEEeqnarray}
\end{theo}
\begin{IEEEproof}
	The proof is provided in Appendix \ref{App6}.
\end{IEEEproof}
The above theorem shows  that, for mirrors, the position closest to   Tx/Rx (Tx) is optimal in the quadratic (linear) power scaling regime because of the adaptive  rotation angle. However, in the saturation regime, since the mirror is large enough to capture all the power,  it achieves the same performance at any point on the ellipse.
\subsubsection{Optimal Position of  DF Relay}
For a DF relay-based link, the optimal position of the relay at high SNR is determined by the diversity gain. Thus,  minimizing the outage performance at high SNR in (\ref{EQ:Opt_prob}) is equivalent to maximizing the diversity gain  of the relay-assisted FSO link, $D_\text{rel}$, and thus, as shown in  \cite{Sahar_Placement}, the  optimal position of the relay $(x_o^*, z_o^*)$  is  equidistant  from the Tx and Rx and  given by 
\begin{IEEEeqnarray}{rll}
	(x_o^*, z_o^*)=\left(0, H_e\right).
	\label{EQ:Opt_pos_rel}
\end{IEEEeqnarray}
\section{Simulation Results}\label{Sec_Sim}

\begin{table}[t]
	\centering
	\caption{System and channel parameters \cite{IRS_FSO_WCNC},  \cite{TCOM_IRSFSO}.}
	\scalebox{0.55}{%
		\begin{tabular}{|| l | c | c || }
			\hline  
			{FSO link Parameters} &{Symbol}& {Value}  \\ 
			\hline
			\hline
			FSO bandwidth &$B_\text{FSO}$ &$1\, \mathrm{GHz}$\\
			FSO wavelength &$\lambda$ &$1550\, \mathrm{nm}$\\
			Beam waist radius &$w_{o1}, w_{o2}$& $2.5\ \mathrm{mm}$\\
			Noise spectral density &$N_0$&$-114\ \mathrm{dBm}/\mathrm{MHz}$\\
			Attenuation coefficient &$\kappa$&$0.43\times 10^{-3} \frac{\mathrm{dB}}{\mathrm{m}}$\\
			Refractive-index structure constant &$C_n^2$ &$50\times 10^{-15}$\\
			Impedance of the propagation medium &$\eta$ &$377 \, \Omega$\\
			\hline
			\hline
			{System Parameters}&&  \\
			\hline
			\hline
			Total transmit power &$P_\text{tot}$ &$0.4\text{W}$\\
			PD  responsivity &$\zeta$ &1\\
			IRS size  &$L_{x}\times L_{y}$ & $1$ m $\times 1$ m\\		
			Lenses radius &$a$&$10\ \mathrm{cm}$\\
			LOS distance between Tx and Rx &$L_\text{tr}$ &$800\,\text{m}$\\
			End-to-end distance  &$d_{3}$ &$1\,\text{km}$\\ 
			Focal distance for QP profile &$f$ & $250\,\text{m}$\\
			IRS and relay positions &$(x_o,y_o,z_o)$ &$(200,0,274.95)\text{m}$\\
			\hline
		\end{tabular}
	}
	\label{Table:Sys_Param}
		\vspace{-8mm}
\end{table}

In the following, we  consider an IRS-assisted FSO link with the parameters given in Table \ref{Table:Sys_Param}, unless  specified otherwise. 
\subsection{Validation of Power Scaling Law }
\begin{figure}[t]
	\centering
	\includegraphics[width=1\textwidth]{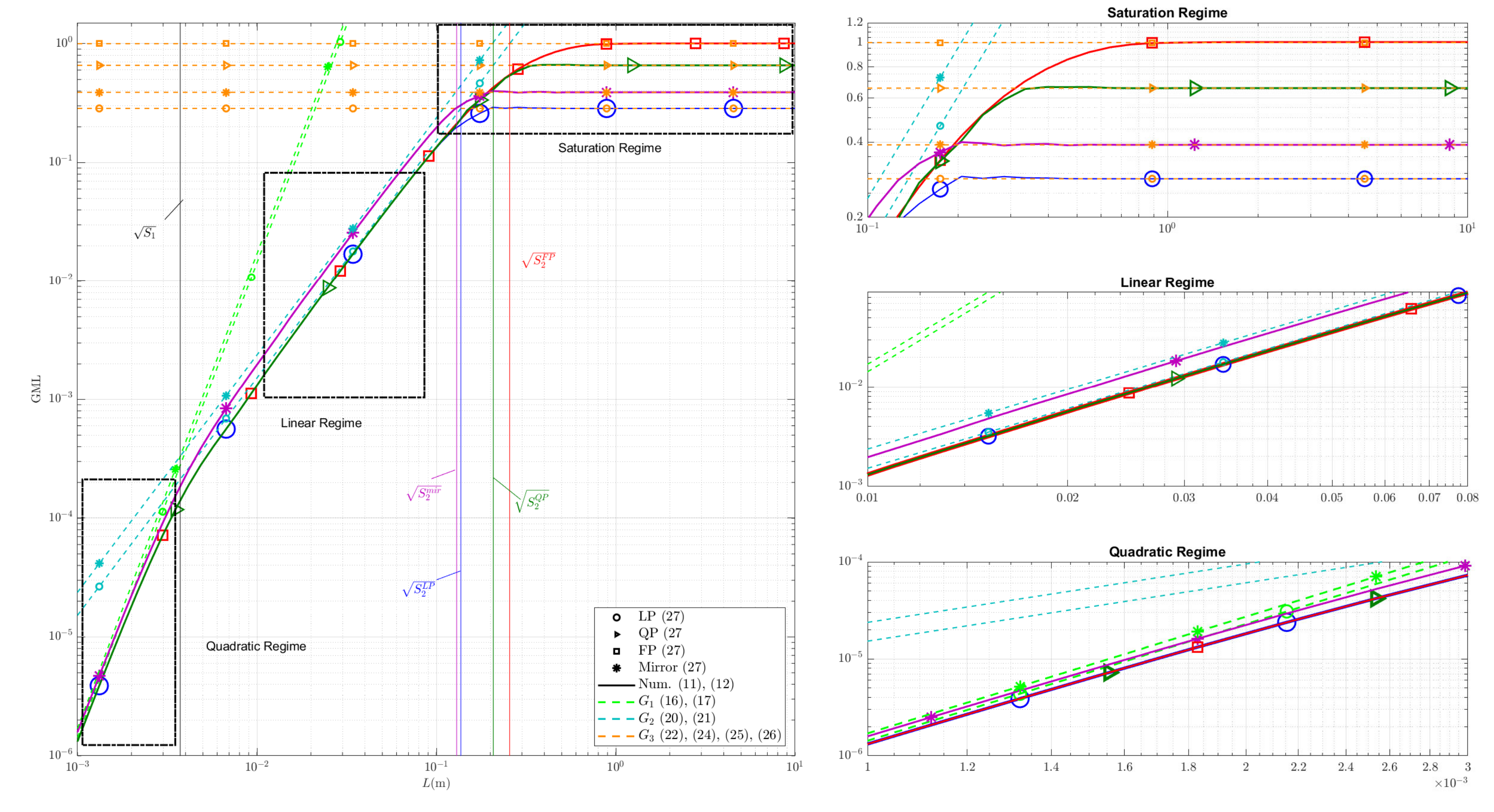}
	\caption { GML, $h_{\text{gml},3}$,  vs. IRS length, $L$, for $w_{o1}=2.5$ mm, $f=\frac{d_3}{4}$, $x_\text{irs}=200$ m.\label{Fig:GML}}
		\vspace{-8mm}
\end{figure}

Fig.~\ref{Fig:GML} shows the GML  of an IRS-assisted FSO link, $h_\text{gml,3}$, versus the length of the square-shaped IRS, $L$, for  mirror-based and MM-based IRSs with different phase shift profiles. As can be observed, the numerical GML  in (\ref{gain}) and (\ref{Huygens}) matches the analytical approximation in (\ref{EQ:Theo1}).  Moreover, depending on the IRS size, the analytical GML in (\ref{EQ:Theo1}) is determined by  ${G}_1^\iota$ in (\ref{EQ:ApproxG_1}), (\ref{EQ:ApproxG_1_mir}), ${G}_2^\iota$ in (\ref{EQ:ApproxG_2}), (\ref{EQ:ApproxG_2_mir}), and $G_3^\iota$  in (\ref{EQ:Lem3}), (\ref{EQ:G3_QP}), (\ref{EQ:Lem_sat_FP}), and (\ref{EQ:Lem3_mir}) for $\iota\in\{\text{LP},\text{QP},\text{FP},\text{mir}\}$. To improve the clarity of the figure, we show $G_1^\iota$ and $G_2^\iota$ only for the LP profile and the mirror-based IRS. Fig.~\ref{Fig:GML} shows  that for  IRS  lengths of $L\leq \sqrt{S_1}$, the approximated GML coefficients $G_1^\iota,\iota\in\{\text{LP},\text{QP},\text{FP}\}$ in (\ref{EQ:ApproxG_1}) and $G_1^\text{mir}$ in (\ref{EQ:ApproxG_1_mir}) coincide with the asymptotic  GML coefficients. Furthermore, for IRS  lengths of $L\leq \sqrt{S_1}$,  the asymptotic GML,  ${G}_1^\iota$, increases quadratically with the IRS size $L^2$, see (\ref{EQ:ApproxG_1}).  Moreover, in this regime, the GML coefficient is identical for  IRSs with LP, QP, and FP profiles, see Lemma \ref{Lemma1}. However, the mirror-based IRS can collect more power as it adjusts its orientation w.r.t. LS and Rx, which results in a slightly higher  GML value.  For IRS lengths in the range  $\sqrt{S_1}\leq L \leq \sqrt{S_2^\iota}$, the IRS collects the power of the tails of the Gaussian beam incident on the IRS and the GML scales linearly with the IRS size $L^2$. The approximated GML coefficient $G_2^\iota$ in (\ref{EQ:ApproxG_2}) and (\ref{EQ:ApproxG_2_mir}) matches well the numerical GML. In this regime, the lens still collects the same amount of power for IRSs with LP, QP, and FP profiles resulting in identical GML coefficients.  
Finally, for IRS lengths of $ L\geq \sqrt{S_2^\iota}$, due to the limited  lens size, the GML coefficient saturates to  $G_3^\iota$ according to (\ref{EQ:Lem3}), (\ref{EQ:G3_QP}), (\ref{EQ:Lem_sat_FP}), and (\ref{EQ:Lem3_mir}) for IRSs with LP, QP, and FP profiles and mirrors, respectively. In this regime, the IRS with the FP profile provides the highest GML coefficient as it can  focus all the power in the lens center. On the other hand, the IRS with LP profile yields  the smallest GML coefficient  since it only reflects the  Gaussian beam which diverges along the propagation path. Moreover, because of its rotation, the mirror-based IRS reflects more power towards  the Rx lens than the MM-based IRS with  LP profile. The GML coefficient for the IRS with QP profile  is larger than that for the  LP profile and    worse than that for the FP profile since for the chosen focal distance $f=\frac{d_3}{4}$,  the beamwidth at the lens is smaller than that for the LP profile and larger than that for the FP profile.  Furthermore, Fig.~\ref{Fig:GML} confirms that the boundary values $S_2^\iota$ are good indicators for  the IRS length at which the transition to the saturation regime occurs. In particular,  IRSs with FP and QP profiles require larger IRS lengths to collect the maximum possible power compared to  IRSs with LP profile and mirrors.

\subsection{Performance Comparison }
\begin{figure}[t]
	\centering
	\begin{subfigure}[h]{0.48\textwidth}
		\centering
		\includegraphics[width=1\textwidth]{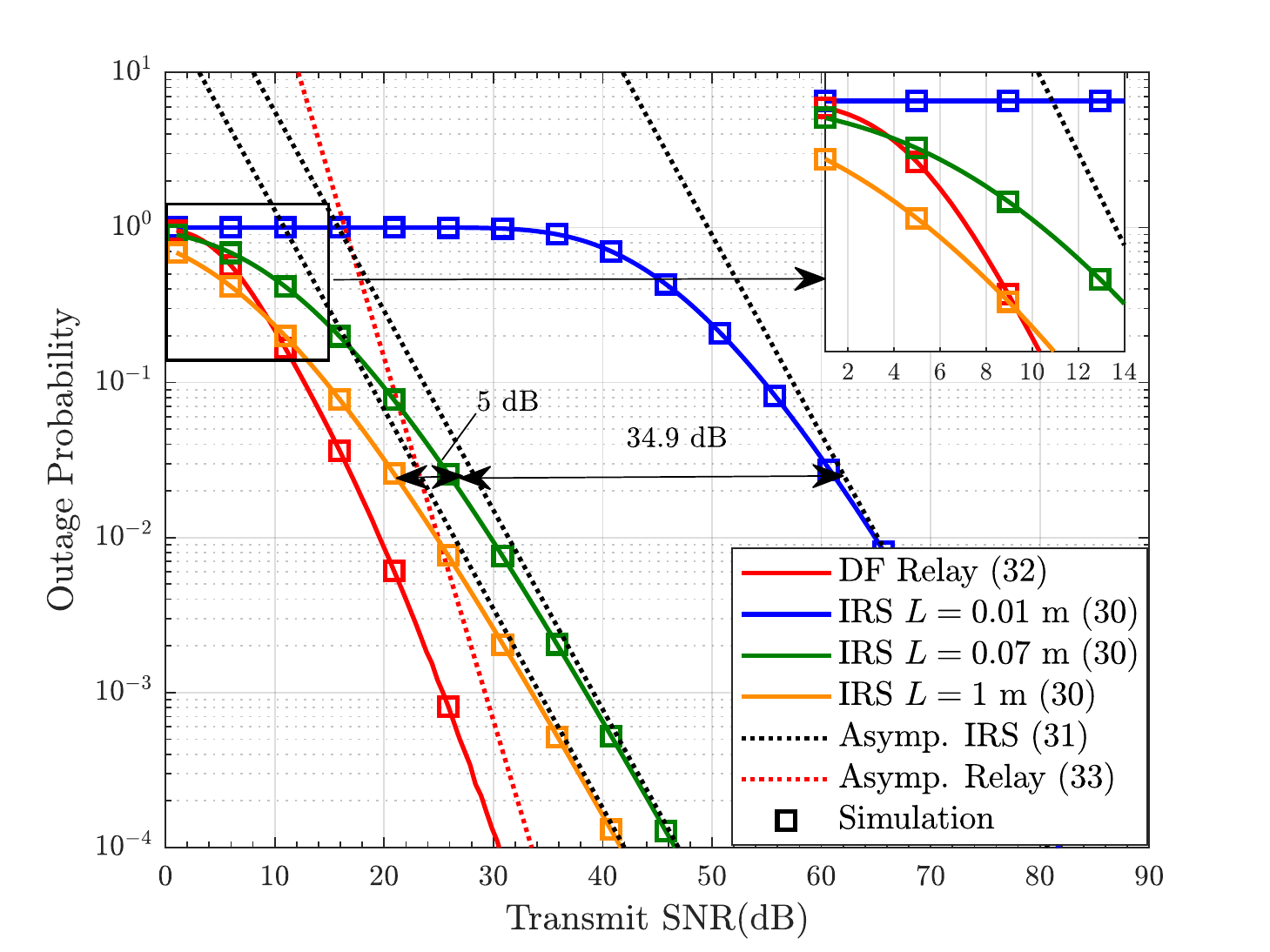}
		\caption{\label{Fig:Outage_Length}} 
	\end{subfigure}
	\begin{subfigure}[h]{0.48\textwidth}
		\centering
		\includegraphics[width=1\textwidth]{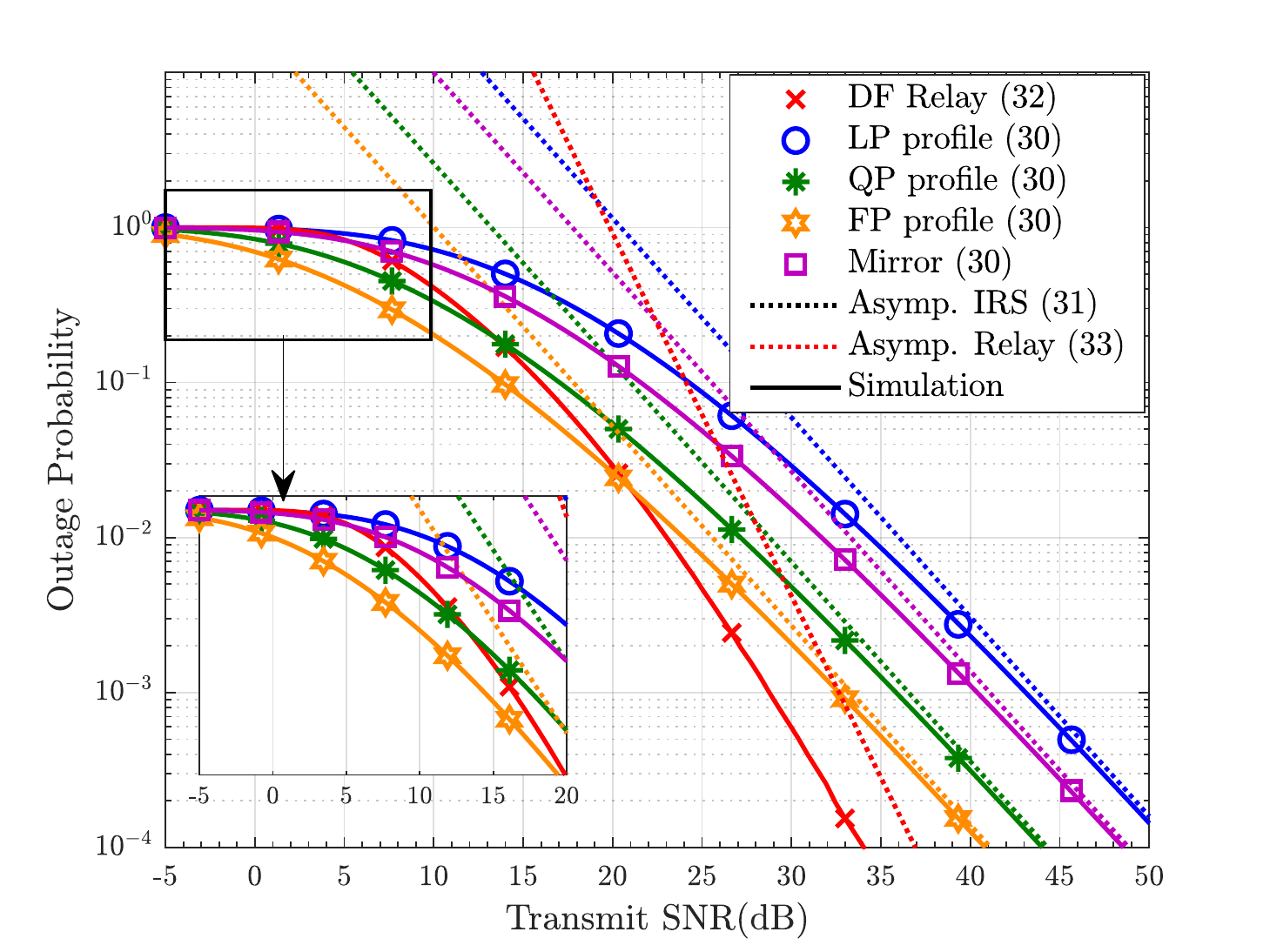}
		\caption{\label{Fig:Outage_Phase}}		
	\end{subfigure}
	\caption {Outage probability vs. transmit SNR a) IRS with LP profile   b) IRS length $L=1$ m with different phase shift profiles. \label{Fig:Outage}}
			\vspace{-8mm}
\end{figure}

Fig.~\ref{Fig:Outage} shows the outage probability  of  relay- and IRS-assisted FSO links for a) IRSs with LP profiles and  lengths  $L=1$ cm, $7$ cm, and 1 m and  b) IRSs with LP, QP, and FP profiles and length $1$ m for a threshold SNR of $\gamma_{th}=0$ dB versus the transmit SNR, $\bar{\gamma}$.   As can be observed, the analytical outage probabilities for the relay in (\ref{EQ:Outage_relay}) and the IRS in (\ref{EQ:Outage_IRS}) match the simulation results.  Furthermore,  the asymptotic outage probabilities  for IRS- and relay-assisted links in (\ref{EQ:Diversity_IRS}) and (\ref{EQ:Diversity_Relay}), respectively, become accurate for high SNR values. Furthermore,  due to distance-dependent fading parameters, the diversity gain of the relay-assisted FSO link  is approximately two times larger than that of  the IRS-assisted link, i.e., $\frac{D_\text{rel}}{D_\text{irs}}=\frac{\min\{\varrho_1,\varrho_2\}}{\varrho_3}=1.9$.  Moreover,  by increasing the IRS length from $1$ cm to $7$ cm, the FSO link gains 34.9 dB in SNR due to the linear scaling of the received power with the IRS size, see Fig.~\ref{Fig:Outage_Length}. However, when the IRS length increases from $7$ cm to $1$ m, the  received power saturates at a constant value and the additional SNR  gain is only 5 dB. Furthermore, Fig.~\ref{Fig:Outage_Length} reveals that for the adopted system parameters, an IRS with LP profile and  $L=1$ m outperforms the relay at low SNR values (SNR$<9$ dB), although the performance difference is  small.
Moreover,  IRSs with FP and LP profiles collect the most and least power at the Rx lens, respectively, see Fig.~\ref{Fig:GML}. Thus, the  IRS-assisted link with the FP profile provides the lowest outage probability  in Fig.~\ref{Fig:Outage_Phase}. Moreover, the IRSs with the QP and FP profiles outperform   the relay for transmit SNR values less than 14 and 20 dB, respectively. However, due to the impact of distance-dependent fading, the relay yields a better performance  compared to the IRS for high SNR values.

\subsection{Optimal Placement}
\begin{figure}[t!]
	\centering
	\begin{subfigure}[h]{0.49\textwidth}
		\centering
		\includegraphics[width=1\textwidth]{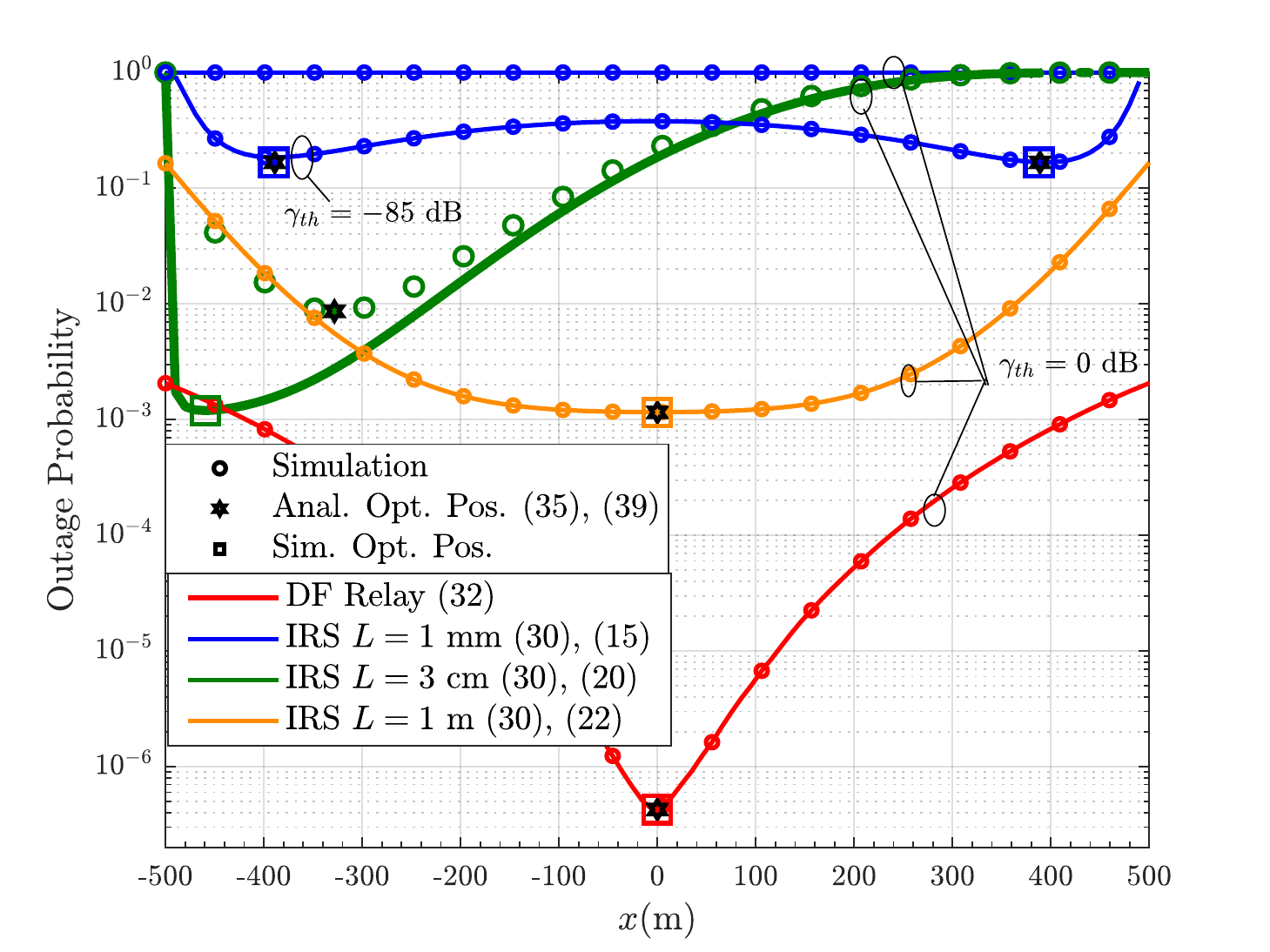}
		\vspace{-1mm}
		\caption{\label{Fig:Pos_Length}} 
	\end{subfigure}
	\begin{subfigure}[h]{0.49\textwidth}
		\centering
		\includegraphics[width=1\textwidth]{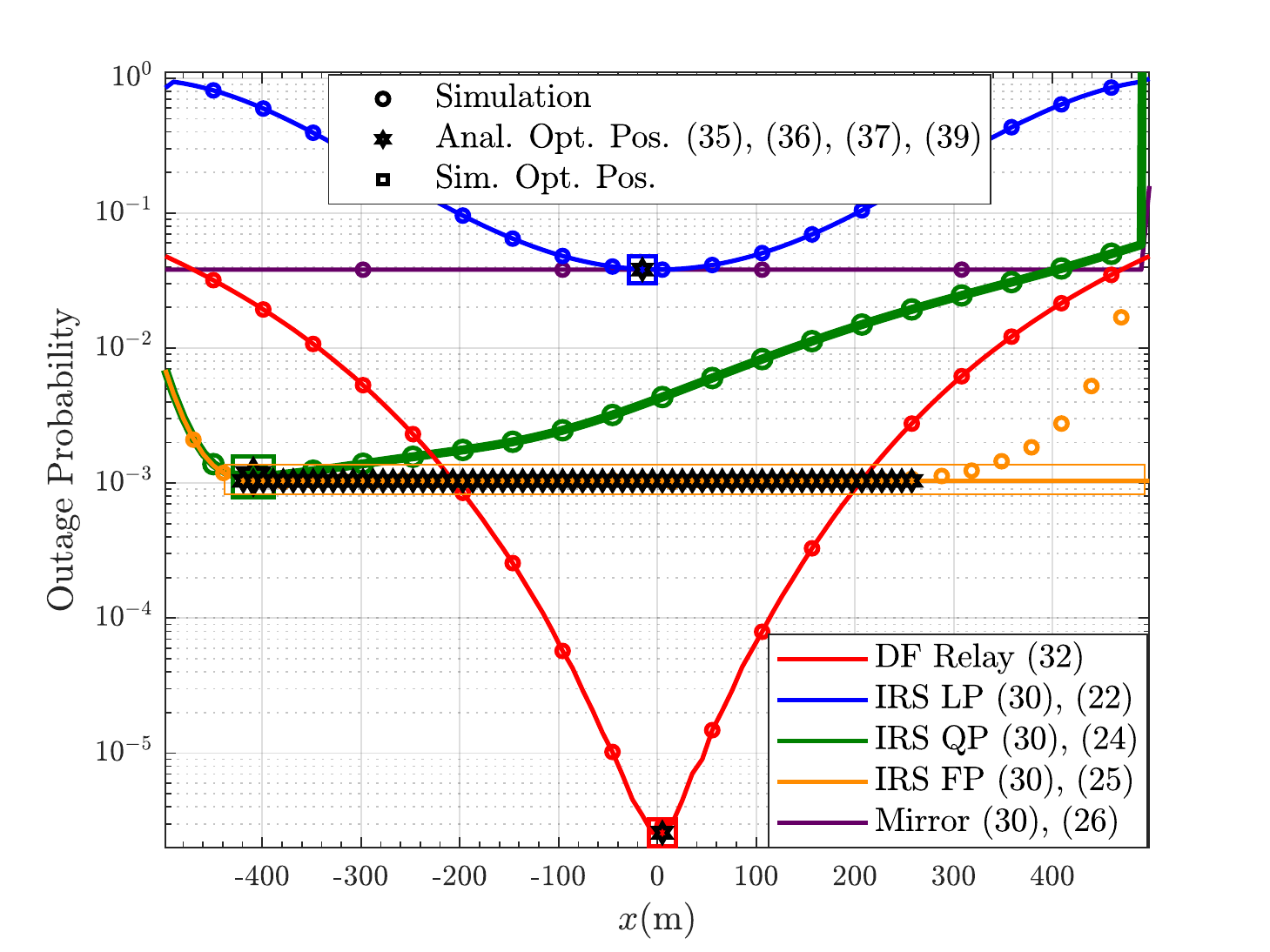}
		\vspace{-1mm}
		\caption{\label{Fig:Pos_Prof}} 		
	\end{subfigure}
	\vspace{-2mm}
	\caption {Outage probability vs. position of the $x$-coordinate of the center of the IRS and the relay, respectively.\label{Fig:Intf} a) LS beam waists $w_{o1}=w_{o2}=7$ mm,  IRS with LP profile and lengths $L=1$ mm, 3 cm, and 1 m. b) LS beam waists $w_{o1}=w_{o2}=2.5$ mm, IRS with length $L=1$ m and   LP, QP ($f=d_3/5$), and FP profiles.\label{Fig:Pos}}
	\vspace{-8mm}
\end{figure}

Fig.~\ref{Fig:Pos_Length} shows the outage probability of the IRS- and relay-assisted links for $\gamma_{th}=0$ dB versus the location of the center of the IRS/relay on the $x$-axis.  To better illustrate the outage performance of extremely small IRSs, we also show results for $\gamma_{th}=-85$ dB. The optimal  positions obtained from the analytical results in (\ref{EQ:theo2}), (\ref{EQ:Opt_pos_rel}) and  simulations are denoted by   \DavidStarSolid\, and $\square$, respectively. The analytical outage probability for   the relay-assisted link in (\ref{EQ:Outage_relay}) matches the simulation results. We obtained the outage probability for the IRS-assisted link in  (\ref{EQ:Outage_IRS})  based on the GMLs $\tilde{G}_1^\iota$ in (\ref{EQ:Lem1}), ${G}_2^\iota$ in (\ref{EQ:ApproxG_2}), and $G_3^\iota$ in (\ref{EQ:Lem3}) for IRS sizes of $L=1$ mm, 3 cm, and 1 m, respectively. The analytical outage performance matches the simulation results except for an IRS length of $L=3$ cm.  The reason is that the IRS with $L=3$ cm does not always operate in the  linear power scaling regime, since the boundary values $S_1$ and $S_2^\text{LP}$ in (\ref{EQ:Theo1}) change with the position of the IRS. However,  despite the  discrepancy between  simulation and analytical results for $x<-200$ m, the analytical optimal placement still leads to a close-to-optimal simulated outage performance. Furthermore, as can be observed, the optimal position of the relay is  equidistant from Tx and Rx  which matches the analytical result in (\ref{EQ:Opt_pos_rel}). Moreover, for  different IRS sizes, different optimal positions are expected.  For a small IRS length of $1$ mm, the IRS operates in the quadratic power scaling regime and the optimal location is close to the Tx or Rx.  However, when the IRS size is large, i.e., $L=1$ m,  the optimal  position is equidistant from Tx and Rx. For IRSs with length $L=3$ cm, the optimal IRS position  is  close to the Tx as expected from (\ref{EQ:theo2}). Furthermore, Fig.~\ref{Fig:Pos_Prof} shows the outage probability performance of mirror-based and MM-based IRSs with LP, QP, and FP profiles for an IRS length  of $L=1$ m and a threshold SNR of $\gamma_{th}=0$ dB. The analytical results were obtained based on (\ref{EQ:Outage_IRS}) and the GML factors $G_3^\text{LP}$ in (\ref{EQ:Lem3}), $G_3^\text{QP}$ in (\ref{EQ:G3_QP}),  $G_3^\text{FP}$ in (\ref{EQ:Lem_sat_FP}), and $G_3^\text{mir}$ in (\ref{EQ:Lem3_mir}).  As can be observed, the mirror-based IRS provides a better performance compared to the IRS with LP profile as the mirror is rotated,  and thus, is able to collect more power.  Furthermore, the  QP profile yields improved performance  by reducing the size of the  beam footprint in the Rx lens plane. Moreover, the FP profile achieves the lowest outage probability for  the IRS-assisted system.  The optimal position of the IRS with LP profile  is  equidistant from Tx and Rx. On the other hand, because of the applied rotation, the  mirror can provide optimal outage performance  regardless  of its position. For the QP profile, the optimal position of the IRS depends on parameter $f$ as shown in (\ref{EQ:Optimal_pos_eq_QP}) and is close to the Tx at $x=-399$ m for the chosen value $f=\frac{d_3}{5}$. For the FP profile,  the analytical outage performance does not always  match  the simulation results. For $x\geq 257 \mathrm{m}$, the IRS size is not large enough to  operate in the saturation regime, and thus, the analytical GML factor $G_3^\iota$ differs from the simulations. Moreover, as analytically shown in  (\ref{EQ:Optimal_pos_eq_FP}), the IRS with FP profile achieves the minimum outage performance  over a wide range of IRS positions, i.e., in the interval  $-416\mathrm{m}\leq x\leq 257 \mathrm{m}$ for the considered case.

\section{Conclusions}\label{Sec_concl}
\vspace*{-1mm}
In this paper, we have  analyzed the power scaling law for IRS-assisted FSO systems. Depending on the beam waist, the position of Tx and Rx w.r.t. the IRS, and the lens radius, the received power at the lens grows  quadratically or linearly with the IRS size or it remains constant. We analyzed the GML,   the boundary IRS sizes, and the asymptotic outage performance for these power scaling regimes.  Our results show that, at the expense of a much higher hardware complexity,  relay-assisted links    outperform  IRS-assisted links at high SNR, but at low SNR, an IRS-assisted link can  achieve superior performance. We also compared the optimal IRS placement for the different power scaling regimes with the optimal relay placement. IRSs operating in the quadratic regime achieve optimal outage   performance close to  Tx and Rx, whereas IRSs operating in linear regime  operate optimally close to Tx. For IRSs in the saturation regime, the optimal placement depends on the IRS phase shift profile. For the LP profile, the IRS performs better when placed equidistant  from  Tx and Rx. For  IRSs with QP profile, the optimal position  can be adjusted via  the focal distance. Moreover,   IRSs with FP profile can achieve their optimal performance over a large range of IRS positions. Finally,  the performance of  mirror-based IRSs is independent from their position due to the rotation of the mirror. 
\appendices
\renewcommand{\thesectiondis}[2]{\Alph{section}:}
\vspace*{-3mm}
\section{Proof of Lemma~\ref{Lemma1}}\label{App1}
\vspace*{-2mm}
First, the electric field of the Gaussian beam in (\ref{Gauss}) incident on the IRS, $E_{\text{in}}({\mathbf{r}_r})$, is given by (\ref{EQ:E_incident})
Then, we adopt $\Phi_\text{irs}(\mathbf{r}_r)= k\left(\Phi_{x^2}x_r^2+\Phi_{y^2}y_r^2+\Phi_{x}x_r+\Phi_{y}y_r+\Phi_0\right)$ for the QP profile in (\ref{EQ:Phase_QP}) and the LP profile in (\ref{EQ:Phase_LP}), where  $\Phi_{x^2}=\Phi_{y^2}=0$ in the latter case.  Next, by
substituting (\ref{EQ:E_incident}) and $\Phi_\text{irs}(\mathbf{r}_r)$  in (\ref{Huygens}) and given that $d_2\gg d_n$, we approximate  ${\lVert\mathbf{r}_o-{\mathbf{r}_r}\rVert}\approx{d_2-{\frac{x_rx_o+y_ry_o}{d_2}}}+\frac{x_r^2+y_r^2}{2d_2}-\frac{x_r^2\cos^2(\theta_r)}{2d_p}$. Thus, we obtain
\begin{IEEEeqnarray}{rll}
	E_{r}\left(\mathbf{r}_{p2}\right)&=\tilde{C}_L \int\limits_{-\frac{L_x}{2}}^{\frac{L_x}{2}}  \exp\left(-jkx_ra_x-x_r^2 a_{x^2}  \right)   \mathrm{d}{x}_r\int\limits_{-\frac{L_y}{2}}^{\frac{L_y}{2}}  \exp\left(-jky_ra_y-y_r^2 a_{y^2}  \right)   \mathrm{d}{y}_r, \qquad\nonumber\\
	&\overset{(a)}{\approx}\tilde{C}_L\int\limits_{-\frac{L_x}{2}}^{\frac{L_x}{2}}  \exp\left(-jkx_r\frac{x_{p2}\sin(\theta_r)}{d_2}  \right)   \mathrm{d}{x}_r\int\limits_{-\frac{L_y}{2}}^{\frac{L_y}{2}}  \exp\left(-jky_r\frac{y_{p2}}{d_2} \right)   \mathrm{d}{y}_r, \qquad\nonumber\\	
	&=\tilde{C}_LL_xL_y\text{sinc}\left(\frac{kL_x\sin(\theta_r)x_{p2}}{2d_2}\right)\text{sinc}\left(\frac{kL_yy_{p2}}{2d_2}\right),
	\label{EQ:Proof_Lem_Quad_LP_I}
\end{IEEEeqnarray}
where $\tilde{C}_L={C}_\ell C_r\sqrt{\sin(\theta_i)} e^{-jk(d_1+d_2)+j\tan^{-1}\left(\frac{d_1}{z_{R1}}\right)-jk\Phi_0}$, $a_{x^2}=\frac{1}{w^2_{\text{in},x}}+jk\Phi_{x^2}+jk\frac{\sin^2(\theta_i)}{2R(d_1)}+\frac{\sin^2(\theta_2)}{2d_2}$, $a_{y^2}=\frac{1}{w^2_{\text{in},y}}+jk\Phi_{y^2}+jk\frac{1}{2R(d_1)}+\frac{1}{2d_2}$, $a_x= -\cos(\theta_{i})+\Phi_x-\frac{x_o}{d_2}$, and $a_y= \Phi_y-\frac{y_o}{d_2}$. In $(a)$, we approximate  the  Gaussian beam incident on the IRS  by a plane wave. This is valid under the following assumptions:
	\begin{itemize}
		\item For the LP profile in (\ref{EQ:Phase_LP}) with $\Phi_{x^2}=\Phi_{y^2}=0$, by assuming $L_x,L_y\ll \sqrt{\frac{2d_1 d_2}{kd_3}}$, $L_x\ll w_{\text{in},x}$ and $L_y\ll w_{\text{in},y}$, we have $-a_{x^2}x_r^2, -a_{y^2}y_r^2\to 0$, which leads to  (\ref{EQ:Proof_Lem_Quad_LP_I}).
		\item For the QP profile in (\ref{EQ:Phase_QP}), by assuming ${L_x,L_y\ll\sqrt{\frac{4f}{k}}}$, $L_x\ll w_{\text{in},x}$, and $L_y\ll w_{\text{in},y}$, we obtain (\ref{EQ:Proof_Lem_Quad_LP_I}).
	\end{itemize}
Moreover, substituting the  FP profile from (\ref{EQ:Phase_FP}) in (\ref{Huygens}), we obtain
\begin{IEEEeqnarray}{rll}
	E_r(\mathbf{r}_{p2})=\tilde{C}_L\iint_{({x}_r,{y}_r)\in\Sigma_\text{irs}} \exp\left(-\frac{x_r^2}{w^2_{\text{in},x}}-\frac{y_r^2}{w^2_{\text{in},y}}\right)e^{-jk\left(\lVert\mathbf{r}_o-\mathbf{r}_r\rVert-\lVert\tilde{\mathbf{r}}_o-\mathbf{r}_r\rVert\right)} \mathrm{d}x_r\mathrm{d}y_r.
	\label{EQ:Proof_LP_i}
\end{IEEEeqnarray}	
Using $\mathbf{r}_o=\mathbf{r}_{p2}\mathbf{R}_\text{rot}+\tilde{\mathbf{r}}_o$, the phase shift profile inside the integral in (\ref{EQ:Proof_LP_i}) can be written as
\begin{IEEEeqnarray}{rll}
	\lVert\mathbf{r}_o-\mathbf{r}_r\rVert-\lVert\tilde{\mathbf{r}}_o-\mathbf{r}_r\rVert&=\lVert\tilde{\mathbf{r}}_o-\mathbf{r}_r\rVert\left[\left(1+\frac{\lVert\mathbf{r}_{p2}\rVert^2}{\lVert\tilde{\mathbf{r}}_o-\mathbf{r}_r\rVert^2}+2\frac{\langle\mathbf{r}_{p2}\mathbf{R}_\text{rot}, \tilde{\mathbf{r}}_o-\mathbf{r}_r\rangle}{\lVert\tilde{\mathbf{r}}_o-\mathbf{r}_r\rVert^2}\right)^{1/2}-1\right]\nonumber\\
	&\overset{(a)}{\approx} \frac{\langle\mathbf{r}_{p2}\mathbf{R}_\text{rot}, \tilde{\mathbf{r}}_o-\mathbf{r}_r\rangle}{\lVert\tilde{\mathbf{r}}_o-\mathbf{r}_r\rVert},
	\label{EQ:Proof_LP_ii}
\end{IEEEeqnarray}	
where $\langle \cdot, \cdot\rangle $ denotes the inner product of two vectors. In $(a)$, given $\frac{\lVert\mathbf{r}_{p2}\rVert}{\lVert\tilde{\mathbf{r}}_o-\mathbf{r}_r\rVert}\ll 1$,  we   use the  Taylor series  approximation  $(1+\varkappa)^{\frac{1}{2}}\approx 1+\frac{\varkappa}{2}$, valid for $\varkappa\ll 1$.  
Assuming $L_x\ll w_{\text{in},x}$ and $L_y\ll w_{\text{in},y}$, we can ignore the term $\exp\left(-\frac{x_r^2}{w^2_{\text{in},x}}-\frac{y_r^2}{w^2_{\text{in},y}}\right)$ in (\ref{EQ:Proof_LP_i}), which leads to the same expression as in (\ref{EQ:Proof_Lem_Quad_LP_I}).
Next, we substitute the reflected electric field in (\ref{EQ:Proof_Lem_Quad_LP_I}), which was shown above to be valid for the LP, QP, and FP profiles, in (\ref{gain}). To solve the resulting integral, we approximate  the circular lens of radius $a$ with a square lens of length $a\sqrt{\pi}$ \cite{Farid_fso}, and   apply  the following  integral result
\begin{IEEEeqnarray}{rll}
	\int \text{sinc}^2(ax) \mathrm{d}x&
	\overset{(a)}{=} \frac{-1}{2a^2x}\left[2ax\text{Si}(2ax)+\cos(2ax)-1\right],
	\label{EQ:SincInt}
\end{IEEEeqnarray}
where in $(a)$, we use the partial integration rule.  This leads to (\ref{EQ:Lem1}) and  completes the proof.
\vspace*{-2mm}
\section{Proof of Lemma~\ref{Lemma2}}\label{App2}
First,  the  electric field received at the lens, $E_r(\mathbf{r}_{p2})$, after the  reflection by the IRS with the LP or QP profiles in (\ref{EQ:Phase_LP}) and (\ref{EQ:Phase_QP}) is given in  \cite[Eq.~(15)]{TCOM_IRSFSO} as follows
\begin{IEEEeqnarray}{rll}
E_r(\mathbf{r}_{p2})= {\frac{\pi{\tilde{C}_L}}{4\sqrt{b_{x,\iota} b_{y,\iota}}}}\, \exp\left({-\frac{k^2}{4b_{x,\iota}d_2}\sin^2(\theta_r)x_{p2}^2-\frac{k^2}{4b_{y,\iota}d_2}{y_{p2}^2}}\right)D(L_x,L_y,\mathbf{r}_{p2}), \iota\in\{\text{LP}, \text{QP}\},\quad
\label{EQ:E_r_QPLP}
\end{IEEEeqnarray}
with 
\begin{IEEEeqnarray}{rll}				 
	D(L_x,L_y,\mathbf{r}_{p2})=&\Bigg[\text{erf}\left({\sqrt{b_{x,\iota}}}\frac{L_x}{2}+\frac{jk\sin(\theta_r)}{2d_2\sqrt{b_{x,\iota}}} x_{p2}\right)-\text{erf}\left({-\sqrt{b_{x,\iota}}}\frac{L_x}{2}+\frac{jk\sin(\theta_r)}{2d_2\sqrt{b_{x,\iota}}}x_{p2}\right)\Bigg]\nonumber\\
	\times&\Bigg[\text{erf}\left({\sqrt{b_{y,\iota}}}\frac{L_y}{2}+\frac{jk}{2d_2\sqrt{b_{y,\iota}}}y_{p2}\right)-\text{erf}\left(-{\sqrt{b_{y,\iota}}}\frac{L_y}{2}+\frac{jk}{2d_2\sqrt{b_{y,\iota}}}y_{p2}\right)\Bigg].
	\label{theo1}
\end{IEEEeqnarray}
Then,  we substitute (\ref{EQ:E_r_QPLP}) in (\ref{gain}) and define $\tilde{x}_p=\frac{x_{p2}\sqrt{2}}{w_{\text{rx},x}^{\iota,\text{g1}}}$ and $\tilde{y}_p=\frac{y_{p2}\sqrt{2}}{w_{\text{rx},y}^{\iota,\text{g1}}}$. Assuming the size of the lens compared to the received beam footprint is large, i.e., $a\gg \min\{ w_{\text{rx},x}^{\iota,\text{g1}},w_{\text{rx},y}^{\iota,\text{g1}}\}$,  we can change   the bounds of the integral  in (\ref{gain}) to $[-\infty, \infty]$  as follows
\begin{IEEEeqnarray}{rll}
	\tilde{G}_2^\iota=\frac{1}{16} &\int_{-\infty}^{\infty}\frac{1}{\sqrt{2\pi}}e^{-\frac{\tilde{x}_p^2}{2}} \left\lvert\text{erf}\left(\zeta_{1,1}+\zeta_{1,2} \tilde{x}_p\right)-\text{erf}\left(-\zeta_{1,1}+\zeta_{1,2} \tilde{x}_p\right)\right\rvert^2 \mathrm{d}\tilde{x}_p\nonumber\\
	\times&\int_{-\infty}^{\infty}\frac{1}{\sqrt{2\pi}}e^{-\frac{\tilde{y}_p^2}{2}} \left\lvert\text{erf}\left(\zeta_{2,1}+\zeta_{2,2} \tilde{y}_p\right)-\text{erf}\left(-\zeta_{2,1}+\zeta_{2,2} \tilde{y}_p\right)\right\rvert^2 \mathrm{d}\tilde{y}_p.
	\label{EQ:Proof_Lem_lin_QP_I}
\end{IEEEeqnarray}
Then, we expand the terms containing the $\text{erf}(\cdot)$ functions in (\ref{EQ:Proof_Lem_lin_QP_I}) using $|\varkappa|^2=\varkappa\varkappa^*$, valid for any complex valued variable $\varkappa$  as follows
\begin{IEEEeqnarray}{rll}
	&\int_{-\infty}^{\infty}\frac{1}{\sqrt{2\pi}}e^{-\frac{\tilde{x}_p^2}{2}} \left\lvert\text{erf}\left(\zeta_{1,1}+\zeta_{1,2} \tilde{x}_p\right)-\text{erf}\left(-\zeta_{1,1}+\zeta_{1,2} \tilde{x}_p\right)\right\rvert^2 \mathrm{d}\tilde{x}_p=\int\limits_{-\infty}^{\infty}\frac{1}{\sqrt{2\pi}}e^{-\frac{\tilde{x}_p^2}{2}} \nonumber\\
	&\times\left[\text{erf}\left(\zeta_{1,1}+\zeta_{1,2} \tilde{x}_p\right)-\text{erf}\left(-\zeta_{1,1}+\zeta_{1,2} \tilde{x}_p\right)\right]\left[\text{erf}\left(\zeta_{1,1}^*+\zeta_{1,2}^* \tilde{x}_p\right)-\text{erf}\left(-\zeta_{1,1}^*+\zeta_{1,2}^* \tilde{x}_p\right)\right] \mathrm{d}\tilde{x}_p\nonumber\\
	&\overset{(a)}{=}	4\Big[\int_{-\infty}^{\infty}\Phi^\prime(\tilde{x}_p)\left|\Phi\left(\sqrt{2}\zeta_{1,1}+\sqrt{2}\zeta_{1,2} \tilde{x}_p\right)\right|^2
	\mathrm{d}\tilde{x}_p+\int_{-\infty}^{\infty}\Phi^\prime(\tilde{x}_p)\left|\Phi\left(-\sqrt{2}\zeta_{1,1}+\sqrt{2}\zeta_{1,2} \tilde{x}_p\right)\right|^2\mathrm{d}\tilde{x}_p\nonumber\\
	&\,\,-\int_{-\infty}^{\infty}\Phi^\prime(\tilde{x}_p)\Phi\left(\sqrt{2}\zeta_{1,1}+\sqrt{2}\zeta_{1,2} \tilde{x}_p\right) \Phi\left(-\sqrt{2}\zeta_{1,1}^*+\sqrt{2}\zeta_{1,2}^* \tilde{x}_p\right)  \mathrm{d}\tilde{x}_p\nonumber\\
	&\,\,-\int_{-\infty}^{\infty}\Phi^\prime(\tilde{x}_p)\Phi\left(-\sqrt{2}\zeta_{1,1}+\sqrt{2}\zeta_{1,2} \tilde{x}_p\right) \Phi\left(\sqrt{2}\zeta_{1,1}^*+\sqrt{2}\zeta_{1,2}^* \tilde{x}_p\right)  \mathrm{d}\tilde{x}_p\Big],
	\label{EQ:Proof_Lem_lin_QP}
\end{IEEEeqnarray}
where in $(a)$,   $\Phi^\prime(x)=\frac{1}{\sqrt{2\pi}}e^{-\frac{x^2}{2}}$ is the univariate normal distribution, and   $\text{erf}(x)=2\Phi(\sqrt{2}x)-1$.
Next, we exploit
\cite[Eq.~(20011)]{Owen_Integ_book}
\begin{IEEEeqnarray}{rll}
	\frac{1}{\sqrt{2\pi}}\int_{-\infty}^{\infty} e^{-\frac{x^2}{2}} \Phi(a+bx)^2 \mathrm{d}x=\Phi(\frac{a}{\sqrt{1+b^2}})-2\text{T}\left(\frac{a}{\sqrt{1+b^2}},\frac{1}{\sqrt{1+2b^2}}\right).
	\label{EQ:Proof_Lem_L_QP_III}
\end{IEEEeqnarray}
Moreover, according to \cite[Eq.~(20010.3)]{Owen_Integ_book}, for any arbitrary $a, b,$ and $c$, we have 
\begin{IEEEeqnarray}{rll}
	&\frac{1}{\sqrt{2\pi}}\int_{-\infty}^{\infty} e^{-\frac{x^2}{2}} \Phi(a+bx)\Phi(c+dx) \mathrm{d}x=\frac{1}{2}\Phi\left(\frac{a}{\sqrt{1+b^2}}\right)+\frac{1}{2}\Phi\left( \frac{c}{\sqrt{1+d^2}}\right)\nonumber\\
	&-\text{T}\left(\frac{a}{\sqrt{1+b^2}},\frac{c+cb^2-abd}{a\sqrt{1+b^2+d^2}}\right)-\text{T}\left(\frac{c}{\sqrt{1+d^2}},\frac{a+ad^2-bcd}{c\sqrt{1+b^2+d^2}}\right)-\begin{cases}
		0&  ac>0,\\
		-\frac{1}{2} & ac<0,
	\end{cases}.
	\label{EQ:Proof_Lem_L_QP_IV}
\end{IEEEeqnarray}
Then, we simplify (\ref{EQ:Proof_Lem_lin_QP}) by exploiting   (\ref{EQ:Proof_Lem_L_QP_III}) and (\ref{EQ:Proof_Lem_L_QP_IV}), and  obtain
\begin{IEEEeqnarray}{rll}
	&\int_{-\infty}^{\infty}\frac{1}{\sqrt{2\pi}}e^{-\frac{\tilde{x}_p^2}{2}} \left\lvert\text{erf}\left(\zeta_{1,1}+\zeta_{1,2} \tilde{x}_p\right)-\text{erf}\left(-\zeta_{1,1}+\zeta_{1,2} \tilde{x}_p\right)\right\rvert^2 \mathrm{d}\tilde{x}_p\nonumber\\
	&\overset{(a)}{=}-8\text{T}(a_{1},c_{1,1})-8\text{T}(a_{1}^*,c_{1,1}^*)+8\text{T}(a_{1},c_{2,1})+8\text{T}(a_{1}^*,c_{2,1}^*)+4.
	\label{EQ:Proof_Lem_lin_QP_V}
\end{IEEEeqnarray}
Moreover, in $(a)$, we used $\Phi(-x)=1-\Phi(x)$, $\text{T}(-a,h)=\text{T}(a,h)$  \cite[pp.~414, Eq.~(2.5)]{Owen_Integ_book}.  Here,
  $c_{1,i}=\frac{\zeta_{1,i}^*\left(1+2\zeta_{2,i}^2-2\frac{\zeta_{1,i}}{\zeta_{1,i}^*}|\zeta_{2,i}|^2\right)}{\zeta_{1,i}\sqrt{1+2\zeta_{2,i}^2+2(\zeta_{2,i}^*)^2}}, i\in\{1,2\}$, can be simplified by substituting $\zeta_{1,1}$ and $\zeta_{1,2}$, which leads to $c_{1,i}=0$. Then, applying $\text{T}(h,0)=0$  \cite[pp.~414, Eq.~(2.1)]{Owen_Integ_book} in (\ref{EQ:Proof_Lem_lin_QP_V}), the result in (\ref{EQ:Proof_Lem_lin_QP_I}) simplifies to (\ref{EQ:Lem2}). We can apply similar  steps for the second integral in (\ref{EQ:Proof_Lem_lin_QP_I}).

Next,  we obtain the  power intensity received by the Rx lens via an IRS with  FP profile, $I_r(\mathbf{r}_{p2})$, by  substituting (\ref{EQ:Proof_LP_ii}) in (\ref{EQ:Proof_LP_i}). Then,  exploiting \cite[Eq.~(2.33-1)]{integral}, we obtain
\begin{IEEEeqnarray}{rll}
	&\int e^{-ax^2-bx}\mathrm{d}x=\frac{1}{2}\sqrt{\frac{\pi}{a}} \exp\left(\frac{b^2}{4a}\right)\text{erf}\left(\sqrt{a}x+\frac{b}{2\sqrt{a}}\right),\quad \forall a\neq 0
	\label{EQ:Proof_Lem_lin_QP_VI}
\end{IEEEeqnarray}
and using $	I_r(\mathbf{r}_{p2})=\frac{1}{2\eta}|E_r(\mathbf{r}_{p2})|^2$, we obtain
\begin{IEEEeqnarray}{rll}
	I_r(\mathbf{r}_{p2})&=C_{r2} \exp\left(-\frac{x_{p2}^2}{(w^{\text{FP},\text{g2}}_{\mathrm{rx},x})^2}-\frac{y_{p2}^2}{(w^{\text{FP},\text{g2}}_{\mathrm{rx},y})^2}\right)\nonumber\\
	&\times \left[\text{erf}\left(\frac{L_x}{2w_{\text{in},x}}+j\frac{x_{p2}}{\sqrt{2}w^{\text{FP},\text{g2}}_{\mathrm{rx},x}}\right)-\text{erf}\left(-\frac{L_x}{2w_{\text{in},x}}+j\frac{x_{p2}}{\sqrt{2}w^{\text{FP},\text{g2}}_{\mathrm{rx},x}}\right)\right]^2\nonumber\\
	&\times \left[\text{erf}\left(\frac{L_y}{2w_{\text{in},y}}+j\frac{y_{p2}}{\sqrt{2}w^{\text{FP},\text{g2}}_{\mathrm{rx},y}}\right)-\text{erf}\left(-\frac{L_y}{2w_{\text{in},y}}+j\frac{y_{p2}}{\sqrt{2}w^{\text{FP},\text{g2}}_{\mathrm{rx},y}}\right)\right]^2,
	\label{EQ:I_FP}
\end{IEEEeqnarray}
where $C_{r2}=\frac{P_\text{tot}\pi w^2(d_1)\sin(\theta_{r})}{8 \lambda^2d_2^2\sin(\theta_{i})}$. Then, we substitute $\tilde{x}_{p}$ and $\tilde{y}_{p}$ and by assuming  $a\gg \min\{ w_{\text{rx},x}^{\text{FP},\text{g2}}, w_{\text{rx},y}^{\text{FP},\text{g2}}\}$, we can replace  the bounds of the integral in (\ref{gain}) by $-\infty$ and $\infty$ as follows
\begin{IEEEeqnarray}{rll}
	G_2^\text{FP}&= \frac{1} {32\pi}\int_{-\infty}^{\infty}e^{-\frac{\tilde{x}^2_{p}}{2}} \left[\text{erf}\left(\zeta_{1,1}+j\frac{\tilde{x}_{p}}{{2}}\right)-\text{erf}\left(-\zeta_{1,1}+j\frac{\tilde{x}_{p}}{{2}}\right)\right]^2\mathrm{d}\tilde{x}_{p}\nonumber\\
	&\times\int_{-\infty}^{\infty}e^{-\frac{\tilde{y}^2_{p}}{2}} \left[\text{erf}\left(\zeta_{2,1}+j\frac{\tilde{y}_{p}}{{2}}\right)-\text{erf}\left(-\zeta_{2,1}+j\frac{\tilde{y}_{p}}{{2}}\right)\right]^2\mathrm{d}\tilde{y}_{p}.
	\label{EQ:Proof_Lem_L_FP}
\end{IEEEeqnarray}
Substituting $\text{erf}(x)=2\Phi(\sqrt{2}x)-1$ in (\ref{EQ:Proof_Lem_L_FP}) and using similar steps as in (\ref{EQ:Proof_Lem_lin_QP_I}), we obtain (\ref{EQ:Lem2}), which completes the proof.
\vspace*{-2mm}
\section{Proof of Lemma~\ref{Lemma3}}\label{App3}
By substituting $L_x,L_y\to \infty$ in $E_r(\mathbf{r}_{p2})$ in (\ref{EQ:E_r_QPLP}), the GML  in (\ref{gain}) becomes
\vspace*{-2mm}

\begin{IEEEeqnarray}{rll}
	&G_3^\text{LP}=C_3\int\limits_{-\frac{a\sqrt{\pi}}{2}}^{\frac{a\sqrt{\pi}}{2}}\!\!\!\!\! e^{-\frac{k^2\sin^2(\theta_p)x_p^2\mathcal{R}\{b_{x,\text{LP}}\}}{2d_2^2|b_{x,\text{LP}}|^2}} \mathrm{d}x_{p2}\times\int\limits_{-\frac{a\sqrt{\pi}}{2}}^{\frac{a\sqrt{\pi}}{2}}   e^{-\frac{k^2 y_{p2}^2\mathcal{R}\{b_{y,\text{LP}}\}}{2d_2^2|b_{y,\text{LP}}|^2}}  \mathrm{d}y_{p2},\qquad
\end{IEEEeqnarray}
where  $C_3=\frac{2P_\text{tot}\sin(\theta_{i})\sin(\theta_r)\pi}{\lambda^2w^2(d_1)d_2^2 |b_{x,\text{LP}}||b_{y,\text{LP}}|}$. Then, substituting   \cite[Eq.~(2.33-2)]{integral}, we obtain (\ref{EQ:Lem3}), which completes the proof.

\vspace*{-2mm}
\section{Proof of Lemma~\ref{Lemma_sat_QP}}\label{App8}
	Applying  (\ref{EQ:Phase_QP}) to  (\ref{EQ:E_r_QPLP}) and substituting $L_x,L_y\to \infty$ in $E_r(\mathbf{r}_{p2})$, the GML  in (\ref{gain}) becomes
\begin{IEEEeqnarray}{rll}
	&G_3^\text{QP}=C_3\int_{-\frac{a\sqrt{\pi}}{2}}^{\frac{a\sqrt{\pi}}{2}}\!\!\!\!\! e^{-\frac{k^2\sin^2(\theta_p)x_{p2}^2\mathcal{R}\{b_{x,\text{QP}}\}}{2d_2^2|b_{x,\text{QP}}|^2}} \mathrm{d}x_{p2}\times\int_{-\frac{a\sqrt{\pi}}{2}}^{\frac{a\sqrt{\pi}}{2}}   e^{-\frac{k^2 y_{p2}^2\mathcal{R}\{b_{y,\text{QP}}\}}{2d_2^2|b_{y,\text{QP}}|^2}}  \mathrm{d}y_{p2},\qquad
\end{IEEEeqnarray}
where  $C_3=\frac{2P_\text{tot}\sin(\theta_{i})\sin(\theta_r)\pi}{\lambda^2w^2(d_1)d_2^2 |b_{x,\text{QP}}||b_{y,\text{QP}}|}$. Then, substituting   \cite[Eq.~(2.33-2)]{integral}, leads to (\ref{EQ:G3_QP}), which completes the proof.

\vspace*{-2mm}
\section{Proof of Lemma~\ref{Lemma_sat_FP}}\label{App11}
Assuming $\rho, \varrho \gg 1$ leads to $\text{erf}(\rho+j\varrho)-\text{erf}(-\rho+j\varrho)\approx 2$ in (\ref{EQ:I_FP}) and thus, we obtain
\begin{IEEEeqnarray}{rll}
	I_r(\mathbf{r}_{p2})&=C_{r2} \times 16 \exp\left(-\frac{2x_{p2}^2}{(w^{\text{FP},\text{g3}}_{\mathrm{rx},x})^2}-\frac{2y_{p2}^2}{(w^{\text{FP},\text{g3}}_{\mathrm{rx},y})^2}\right).
	\label{EQ:Proof_Lem_sat_FP_I}
\end{IEEEeqnarray}	
 Then, by substituting (\ref{EQ:Proof_Lem_sat_FP_I}) in  (\ref{gain}) and using \cite[Eq.~(2.33-2)]{integral}, we obtain (\ref{EQ:Lem_sat_FP}), which completes the proof.
\vspace*{-2mm}
\section{Proof of Theorem~\ref{Lemma5}}\label{App5}
Depending on the IRS size, the optimal position of the IRS is calculated by approximating $h_{\text{gml},3}$ for the respective power scaling regime. First, the position of the center of the IRS $(x_o,z_o)$ on the ellipse can be rewritten in terms of $d_1$ and $d_2$ as  follows
\vspace*{-2mm}
\begin{IEEEeqnarray}{rll}
	x_o=\frac{d_1^2-d_2^2}{2L_\text{tr}}, z_o=H_e\Big[1-\frac{\left(d_1^2-d_2^2\right)^2}{d_{3}^2 L_\text{tr}^2}\Big]^{1/2}.\quad
	\label{proof_Theo2_0}
\end{IEEEeqnarray}
For $\Sigma_\text{irs}\leq \min\{S_1, S_3^\iota\} $, the  GML  is $h_{\text{gml},3}\approx{G}_1$. Then, we substitute in (\ref{EQ:ApproxG_1}) the values of $\sin(\theta_i)=\frac{z_o}{d_1}$,  $\sin(\theta_p)=\frac{z_o}{d_2}$,  $z_o$ given in (\ref{proof_Theo2_0}),  and $d_2=d_3-d_1$.
Next, by solving $\frac{\partial{G}_1}{\partial d_1}=0$, the extremal points, comprising maxima and minima, are given by
\begin{IEEEeqnarray}{rll}
	&d_{1,\text{min}}^{(1)}= \frac{d_{3}}{2},\,	d_{1, \text{max}}^{(2)}=\frac{d_{3}}{2}+ \frac{\sqrt{2\rho_1}}{4},\,
	&d_{1,\text{max}}^{(3)}=\frac{d_{3}}{2}- \frac{\sqrt{2\rho_1}}{4}.\qquad
	\label{proof_Theo2_2}
\end{IEEEeqnarray}
Then,   substituting the maxima in (\ref{proof_Theo2_0})  leads to (\ref{EQ:theo2}).

Next, for $S_1\leq \Sigma_\text{irs}\leq S_2$, the GML  is $h_{\text{gml},3}\approx{G}_2$. Then, we substitute $\sin(\theta_i)=\frac{z}{d_1}$ and $d_2=d_{3}-d_1$ in (\ref{EQ:ApproxG_2}).
Then, by solving $\frac{\partial{G}_2}{\partial d_1}=0$, the extremal points are obtained as
\vspace*{-2mm}
\begin{IEEEeqnarray}{rll}
	d_{1,\text{max}}^{(1)}=\left(5d_{3}+ \sqrt{\rho_2}\right)/8,\quad
	d_{1,\text{max}}^{(2)}=\left(5d_{3}- \sqrt{\rho_2}\right)/8.
\end{IEEEeqnarray}
Here, $d_{1,\text{max}}^{(1)}$ does not lie on  the ellipse, since  $d_{1,\text{max}}^{(1)}>\max(d_1)$, where $\max(d_1)=\frac{d_{3}+L_\text{tr}}{2}$.  Then, substituting $d_{1,\text{max}}^{(2)}$ in (\ref{proof_Theo2_0}) leads to (\ref{EQ:theo2}).

Next, for $\Sigma_\text{irs}>\max\{S_2, S_3^\iota\} $, the GML  is $h_{\text{gml},3}\approx G_3$. Then, assuming $d_1\gg z_{R1}$, we  substitute $w(d_1)\approx\frac{\lambda d_1}{\pi w_{o1}}$, $R(d_1)=d_1$, and $d_2=d_{3}-d_1$  in (\ref{EQ:Lem3})	 as follows
\begin{IEEEeqnarray}{rll}
G_3^\text{LP}=\underbrace{\text{erf}\left(\sqrt{\frac{\pi}{2}} \frac{a}{w^{\text{LP},\text{g3}}_{\text{rx},x}(d_1)}\right)}_{=G_{3a}^\text{LP}}\underbrace{\text{erf}\left(\sqrt{\frac{\pi}{2}} \frac{a}{w^{\text{LP},\text{g3}}_{\text{rx},y}(d_1)}\right)}_{=G_{3b}^\text{LP}},
\label{EQ:proofa_Lamma6}
\end{IEEEeqnarray}
where  the equivalent beamwidths are $w^{\text{LP},\text{g3}}_{\text{rx}, x}(d_1)= \frac{\lambda(d_3-d_1)^2}{\pi w_{o1} d_1}\left[{\left(\frac{\pi w_{o1}^2}{\lambda d_1}\right)^2+\left(1+\frac{d_1^3}{(d_3-d_1)^3}\right)^2}\right]^{1/2}$ and  $w^{\text{LP},\text{g3}}_{\text{rx}, y}(d_1)=\frac{\lambda d_1}{\pi w_{o1}}  \left[{\left(\frac{\pi w_{o1}^2(d_3-d_1)}{\lambda d_1^2}\right)^2+\left(\frac{d_3}{d_1}\right)^2}\right]^{1/2}$.
The maxima of the $\text{erf}(\cdot)$ functions in (\ref{EQ:Lem3}) occur for the minimum of the beamwidths $w^{\text{LP},\text{g3}}_{\text{rx}, x}$ and $w_{\text{rx}, y}^{\text{LP},\text{g3}}$, which are both convex w.r.t. $d_1$. By solving  $\frac{\mathrm{d} w^{\text{LP},\text{g3}}_{\text{rx}, i}}{\mathrm{d}d_1}=0, i\in\{x,y\}$, we obtain  minimal points at $\frac{d_{3}}{2}$. Thus, both $\text{erf}(\cdot)$ functions in (\ref{EQ:Lem3})  are maximized at $\frac{d_{3}}{2}$, which in turn maximizes $G_3^\text{LP}$. Substituting $\frac{d_{3}}{2}$ in (\ref{proof_Theo2_0}) leads to (\ref{EQ:theo2}).

Next, to determine the optimal placement of an IRS with QP profile, we follow the same steps  as in (\ref{EQ:proofa_Lamma6}) and  obtain
\begin{IEEEeqnarray}{rll}
	G_3^\text{QP}=\underbrace{\text{erf}\left(\sqrt{\frac{\pi}{2}} \frac{a}{w_{\text{rx},x}^{\text{QP},\text{g3}}(d_1)}\right)}_{=G_{3a}^\text{QP}}\underbrace{\text{erf}\left(\sqrt{\frac{\pi}{2}} \frac{a}{w_{\text{rx},y}^{\text{QP},\text{g3}}(d_1)}\right)}_{=G_{3b}^\text{QP}},
	\label{EQ:proofa_Lamma7}
\end{IEEEeqnarray}
where $w_{\text{rx},x}^{\text{QP},\text{g3}}(d_1)=\frac{\lambda}{\pi w_{o1}} \sqrt{z_{R1}^2\frac{(d_3-d_1)^4}{d_1^4}+\frac{d_1^4}{4f^2}}$ and $w_{\text{rx},y}^{\text{QP},\text{g3}}(d_1)=\frac{\lambda}{\pi w_{o1}} \sqrt{z_{R1}^2\frac{(d_3-d_1)^2}{d_1^2}+\frac{(d_3-d_1)^2d_1^2}{4f^2}}$. Here, $w_{\text{rx},y}^{\text{QP},\text{g3}}(d_1)$ is a concave and increasing function, whereas $w_{\text{rx},x}^{\text{QP},\text{g3}}(d_1)$ is a convex function w.r.t. $d_1$, and thus, $G_3^\text{QP}$ is a concave function. Thus, to find the optimal solution, we obtain $\frac{\partial G_3^\text{QP}}{\partial d_1}=0$ as follows
\begin{IEEEeqnarray}{rll}
	\frac{\partial G_3^\text{QP}}{\partial d_1}=\frac{\partial G_{3a}^\text{QP}}{\partial w_{\text{rx},x}^{\text{QP},\text{g3}}}\frac{\partial w_{\text{rx},x}^{\text{QP},\text{g3}}}{\partial d_1}G_{3b}^\text{QP}+\frac{\partial G_{3b}^\text{QP}}{\partial w_{\text{rx},y}^{\text{QP},\text{g3}}}\frac{\partial w_{\text{rx},y}^{\text{QP},\text{g3}}}{\partial d_1}G_{3a}^\text{QP}=0
	\label{EQ:proofa_Lamma8}
\end{IEEEeqnarray}
Using  $\frac{\partial \text{erf}(z)}{\partial z}=\frac{2}{\sqrt{\pi}}e^{-z^2}$ and substituting $\omega_1(d_1)=w_{\text{rx},x}^{\text{QP},\text{g3}}(d_1)$ and $\omega_2(d_1)=w_{\text{rx},y}^{\text{QP},\text{g3}}(d_1)$, we obtain the optimal $d_1^\text{QP}$ by solving (\ref{EQ:Optimal_pos_eq_QP}).

Next, to determine the optimal position of the IRS with FP profile, we substitute $w_{\text{rx},x}^{\text{FP},\text{g3}}(d_1)=w_0\frac{(d_3-d_1)^2}{d_1^2}$  and $w_{\text{rx},x}^{\text{FP},\text{g3}}(d_1)=w_0\frac{d_3-d_1}{d_1}$ in (\ref{EQ:Lem_sat_FP}). Then, following the same steps as for the QP profile, we find the optimal $d_1^\text{FP}$ as the solution of (\ref{EQ:Optimal_pos_eq_FP}) and this completes the proof.
 
 \section{Proof of Theorem~\ref{Lemma6}}\label{App6}
The  placement of the mirror in the quadratic regime can be optimized by substituting $\theta_i^\text{mir}=\frac{\theta_i+\theta_r}{2}$ in (\ref{EQ:ApproxG_1_mir}), and we obtain 
\begin{IEEEeqnarray}{rll}
	 G_1^\text{mir}(d_1)=\frac{G_0}{d_1^2(d_3-d_1)^2} \sin^2\left(\frac{\theta_{i}+\theta_r}{2}\right)\overset{(a)}{=} \frac{d_3^2-L_\text{tr}^2}{4d_1^3(d_3-d_1)^3},
	\label{EQ:proofa_Theo2}
\end{IEEEeqnarray}
 where $G_0=\frac{2\pi^2 w_{o1}^2a^2\Sigma_\text{irs}^2}{\lambda^4}$ and in $(a)$, we apply $\sin^2(x)=\frac{1-\cos(2x)}{2}$ and the cosine rule $\cos(\theta_i+\theta_r)=\frac{L_\text{tr}^2-d_1^2-d_2^2}{2d_1d_2}$. Then, since $G_1^\text{mir}(d_1)$ is a convex function, applying $\frac{\partial G_1^\text{mir}}{\partial d_1}=0$  leads to one minimal point at $d_{1,\text{min}}^{(1)}=\frac{d_3}{2}$ and we can consider   $d_{1,\text{max}}^{(2)}=\max(d_1)=\frac{d_3+L_\text{tr}}{2}$ and $d_{1,\text{max}}^{(3)}=\min(d_1)=\frac{d_3-L_\text{tr}}{2}$ as the maximal points. 
 
 Next, for the linear regime, we can apply similar steps as in (\ref{EQ:proofa_Theo2}) to (\ref{EQ:ApproxG_2_mir}) and obtain $G_2^\text{mir}(d_1)=\frac{d_3^2-L_\text{tr}^2}{2d_1^2d_2}$ as a monotonically decreasing function. This means that $d_{1,\text{max}}^{(1)}=\min(d_1)=\frac{d_3-L_\text{tr}}{2}$ leads to the  maximal value of $G_2^\text{mir}$.
 
Finally, for the  saturation regime , we substitute $w(d_1)\approx \frac{\lambda d_1}{\pi w_{o1}}$ in (\ref{EQ:Lem3_mir}) and  obtain $w_\text{rx}^{\text{mir},\text{g3}}(d_1)\approx \frac{\lambda d_2}{\pi w_{o1}} \sqrt{\frac{z_{R1}^2}{d_1^2}+\left(1+\frac{d_1}{d_2}\right)^2}$. Since $z_{R1}\ll d_1$, then, $w_\text{rx}^{\text{mir},\text{g3}}\approx \frac{\lambda d_3}{\pi w_{o1}}$ and thus, the received beamwidth of the mirror-assisted link does not depend on the position of the mirror. Then, we substitute the derived optimal values for $d_1$ in (\ref{proof_Theo2_0}). This leads to (\ref{EQ:theo3}) and  completes the proof.

\bibliographystyle{IEEEtran}
\bibliography{My_Citation_1-07-2022}
\end{document}